\documentclass[letterpaper, pra,twocolumn, american, aps, eqsecnum, superscriptaddress]{revtex4}
\usepackage{amsfonts}%
\usepackage{amsmath}%
\usepackage{amssymb}%
\usepackage{graphicx}
\usepackage{color}
\usepackage{lipsum}
\usepackage{mathtools}
\providecommand{\U}[1]{\protect\rule{.1in}{.1in}}
\newtheorem{theorem}{Theorem}

\newtheorem{corollary}[theorem]{Corollary}

\newtheorem{definition}[theorem]{Definition}

\newtheorem{lemma}[theorem]{Lemma}

\newtheorem{proposition}[theorem]{Proposition}

\newenvironment{proof}[1][Proof]{\noindent\textbf{#1.} }{\ \rule{0.5em}{0.5em}}

\newcommand{\bes} {\begin{subequations}}
\newcommand{\ees} {\end{subequations}}
\newcommand{\bea} {\begin{eqnarray}}
\newcommand{\eea} {\end{eqnarray}}
\newcommand{\beq}{\begin{equation}}
\newcommand{\eeq}{\end{equation}}

\def\>{\rangle}
\def\<{\langle}
\def\Tr{\textrm{Tr}}

\renewcommand{\max}{\textrm{max}}

\newcommand{\ignore}[1]{}

\definecolor{nblue}{rgb}{0.2,0.2,0.7}
\definecolor{ngreen}{rgb}{0.2,0.6,0.2}
\definecolor{nred}{rgb}{0.7,0.2,0.2}
\definecolor{nblack}{rgb}{0,0,0}

\begin{document}

\title{Symmetry-Protected Topological Entanglement}

\author{Iman Marvian}
\affiliation{Research Laboratory of Electronics, Massachusetts Institute of Technology, Cambridge, MA 02139}
\affiliation{Department of Physics and Astronomy, Center for Quantum Information Science and Technology, University of Southern California, Los Angeles, CA 90089}

\begin{abstract}
We propose an order parameter for the Symmetry-Protected Topological (SPT) phases which are protected by Abelian on-site symmetries. This order parameter, called the \emph{SPT-entanglement}, is defined as the entanglement between $A$ and $B$, two distant regions of the system, given that the total charge (associated with the symmetry) in a third region $C$ is measured and known, where $C$ is a connected region surrounded by $A$, $B$ and the boundaries of the system. In the case of 1-dimensional systems we prove that in the limit where $A$ and $B$ are large and far from each other compared to the correlation length, the SPT-entanglement remains constant throughout a SPT phase, and furthermore, it is zero for the trivial phase while it is nonzero for all the non-trivial phases.   Moreover, we show that the SPT-entanglement is invariant under the low-depth quantum circuits which respect the symmetry, and hence it remains constant throughout a SPT phase in the higher dimensions as well. Also, we show that there is an intriguing connection between SPT-entanglement and the Fourier transform of the string order parameters, which are the traditional tool for detecting SPT phases. This leads to a new  algorithm for extracting the relevant information about the SPT phase of the system from the string order parameters. Finally, we discuss implications of our results in the context of measurement-based quantum computation.



\end{abstract}
\maketitle
\tableofcontents
\newpage
\section{Introduction}

Symmetry-Protected Topological (SPT) order  is a new kind of order at  zero temperature which cannot be described by the traditional Landau paradigm of symmetry breaking \cite{Clas-Wen,Clas-Wen2,Pollman-Turner2, Schuch-Cirac}. Different SPT orders remain distinct from each other in the presence of an appropriate symmetry. However,  there is no local order parameter to distinguish SPT ordered phases from the  trivial phase. Also,  SPT ordered phases  do not  exhibit long-range order between distant sites.   First known example of a SPT phase is the Haldane phase of the antiferromagnetic spin chains with odd integer spins  \cite{AKLT, Hidden-Sym}, which has been recently realized in several experiments  (See e.g. \cite{Exp1, Exp2, Exp3, Exp4}). This phase is protected by the $D_2\simeq \mathbb{Z}_{2}\times\mathbb{Z}_{2}$ symmetry corresponding to $\pi$ rotations around a set of orthogonal axis.  Recently  SPT order has attracted a considerable amount of attention in quantum information community (See e.g. \cite{ElseBartlettDoherty, Jac, Else2012, Henry, Robert, Yoshida}). 


It is often said that (symmetry-protected) topological order is related to the entanglement structure of state.  Hence, it is desirable to have order parameters which directly detect this entanglement structure. In fact, in the case of intrinsic topological order, topological entanglement entropy can reveal the presence of topological order based on the entanglement properties of the ground state \cite{Paolo, Kitaev, Lev-Wen}. 

In this paper, we propose a new quantity, called \emph{SPT-Entanglement}, which detects SPT ordered phases based on their entanglement properties. We show that  for Abelian symmetries and in the appropriate limits,  SPT-Entanglement is a \emph{universal} quantity in all dimensions,  that is it remains constant throughout a SPT phase, similar to the fact that topological entanglement entropy remains constant throughout a topological phase  \cite{ Kitaev, Lev-Wen}.  Furthermore,  in the case of 1-dimensional systems  we calculate SPT-Entanglement in the  Matrix Product State (MPS) framework,  and show that it always successfully detects the presence of SPT order.  We also show that there is an intriguing  connection between the concept of SPT-Entanglement and  the string order parameters. More precisely, we show that SPT-Entanglement can be expressed in terms of the Fourier transform of the string order parameters. 

Although the main results of this paper are focused on the field of SPT order, it also contributes by showing that the \emph{resource theory} point of view to entanglement can be useful in the study of many-body systems (See e.g. \cite{Resource, Horodecki} for recent reviews).  In this point of view, rather than quantifying entanglement using a particular measure of entanglement,  entanglement of a given state is characterized in terms of the  equivalence class of all states that can be reversibly reached from this state via Local Operations and Classical Communication (LOCC). This point of view to entanglement is essential for understating and proving  the properties of SPT-Entanglement. 
This suggests that the resource theory point of view to entanglement might be useful in other many-body problems. 

This point of view also enables us to clearly see the advantage of using entanglement measures instead of correlation measures such as mutual information.  Although in the recent years entanglement measures have been extensively used in the study of many-body systems, it is not always clear that why they are relevant in this context, and why they cannot be replaced by other measures of correlation, such as mutual information, which are often easier to calculate. Interestingly, in the arguments presented in this paper, one can clearly see that the monotonicity of measures of entanglement under classical communication, which distinguishes them from general measures of correlations, is crucial for having a universal quantity which remains constant throughout a SPT phase.

The organisation of the paper is as follows. We start by reviewing some preliminary concepts. In particular,  in Sec.(\ref{Sec:charge}) we review the notions of symmetry and charge, and we discuss the example of $\mathbb{Z}_2\times \mathbb{Z}_2$ symmetry. Then, in Sec.(\ref{Sec:low-depth}) we briefly review the notion of low-depth symmetric circuits. Next, in Sec.(\ref{Sec:Per:Ent}) we review some basic concepts of entanglement theory, namely the concepts of measures  of entanglement and LOCC protocols. Then, in Sec.(\ref{SPT-entanglement}) we define the notion of SPT-entanglement, and present a summary of our main results.  In Sec.(\ref{Sec:strin}) we show that there is a connection between the notion of SPT-Entanglement and the string order parameters. Next, in Sec.(\ref{Sec:Example}) we consider the example of perturbed cluster Hamiltonian, and calculate the SPT-Entanglement for its ground state, which is in the SPT phase protected by $\mathbb{Z}_2\times \mathbb{Z}_2$ symmetry.  Then, in Sec.(\ref{Sec:1-d}) we calculate the SPT entanglement of one-dimensional systems using the MPS classification of SPT phases.   Finally, in Sec.(\ref{Sec:D}) we present the formal versions of our general results on SPT-Entanglement, based on the notion of low-depth symmetric circuits.

\section{Preliminaries}\label{Sec:pre}
\subsection{From symmetry to charge}\label{Sec:charge}

Consider a finite-dimensional lattice system. Let $G$ be an Abelian symmetry which has a unitary on-site representation on this lattice. In particular, let   $u_l(g)$ be the (linear) unitary representation of the group element $g\in G$ on site $l$, such that $u_l(g_1) u_l(g_2)=u_l(g_2) u_l(g_1)=u_l(g_1g_2)$ for all $g_1,g_2\in G$. Note that the symmetry $G$ can be an Abelian subgroup of a non-Abelian symmetry of the Hamiltonian.

 Let $C$ be an arbitrary subset of sites on this lattice. Then, the action of the group element $g\in G$ on $C$ is represented  by the unitary $U^{(C)}(g)=\bigotimes_{l\in C} u_l(g)$.  Since all unitary  Irreducible Representations (irreps) of an Abelian group  are 1-dimensional, this representation can be decomposed as
\beq\label{dec}
U^{(C)}(g)=\bigotimes_{l\in C} u_l(g) = \sum_{\kappa} e^{i\kappa(g)} \Pi_{\kappa}^{(C)}\ ,
\eeq
where each $ e^{i\kappa(g)}$ is a 1-dimensional representation (character) of the group $G$,  and  $\Pi_{\kappa}^{(C)}$ is the projector to the corresponding subspace. 

As we will see in the following sections, instead of thinking in terms of 1-dimensional irreps $e^{i\kappa(g)}$,  it is sometimes more useful to consider  the functions $\kappa(g)$ corresponding to the arguments of the 1-dimensional irreps, and interpret them as different \emph{charges}. Therefore, in the above decomposition each projector $\Pi^{(C)}_{\kappa}$ can be interpreted as the projector to the sector with charge $\kappa$ in region $C$. Each charge $\kappa$ is a real function over the group $G$. Note that the charges are defined only up to a $2\pi$ shift.
 The number of distinct irreps of an Abelian group $G$ are equal to $|G|$,  the order of group. Therefore, the set of distinct charges, denoted by $Q$, has also  $|G|$ different elements. 
Note that this set itself forms an additive group. That is for any charges $\kappa_1, \kappa_2\in Q$,  we have
\beq
\kappa_1+\kappa_2=\kappa_3\  (\text{mod\ } 2\pi)\ ,
\eeq
for some $\kappa_3\in Q$.   

Recall the orthogonality relations between  1-dimensional irreps  of Abelian groups, 
\beq
\sum_{g\in G} e^{i\kappa_1(g)} e^{-i\kappa_2(g)}=|G|\times \delta_{\kappa_1,\kappa_2}\ ,
\eeq
where $\delta$ is the Kronecher delta function. Using these relations, we find that the projector to the subspace with charge $\kappa$ can be written as 
\beq\label{aero}
\Pi_{\kappa}^{(C)}=\frac{1}{|G|}\sum_{g\in G} e^{-i\kappa(g)} U^{(C)}(g) \ .
\eeq
It can be easily shown that  the total charge in a region can be written as the sum of the charges in the subregions of this region. In particular, if we partition $C$ into two non-overlapping subregions $C_1$ and $C_2$ such that $C=C_1\cup C_2$, then  
\beq\label{addit}
\Pi_{\kappa}^{(C)}= \sum_{\substack{\kappa_1,\kappa_2\in Q\\
                  \kappa_1+\kappa_2=\kappa\ \text{(mod}\ 2\pi) }} \Pi_{\kappa_1}^{(C_1)}\otimes \Pi_{\kappa_2}^{(C_2)}\ , 
\eeq
where the summation is over all charges $\kappa_1,\kappa_2\in Q$ which add up to $\kappa$ (mode $2\pi$). Note that the charges corresponding to non-Abelian groups are not additive.

Let 
\beq
U(g)=\bigotimes_{l\in C} u_l(g)=\sum_{\kappa\in Q} e^{i\kappa(g)} \Pi_\kappa
\eeq
 be the representation of the group element $g\in G$, on all sites in the system, and $\Pi_\kappa$ be the projector to the subspace with the total charge $\kappa$ in the system. We call a unitary transformation $V$ \emph{symmetric} if it commutes with  the action of the symmetry group, that is
\beq
\forall g\in G:\ \ U(g) V U^\dag(g)=V\  ,
\eeq
or equivalently, $[V,U(g)]=0$. This definition implies
\beq
\forall \kappa\in Q:\ \  V\Pi_\kappa V^\dag=\Pi_\kappa\ ,
\eeq
which basically means that the total charge in the system is conserved under symmetric unitaries. 

As we will see in the rest of the paper, this conservation law, together with the additivity relation in Eq.(\ref{addit}), which holds only for  Abelian groups, play important roles in our arguments. Indeed, a useful advantage of thinking in terms of  additive charges instead of 1-dimensional irreps, is that one can use the standard intuition about the additivity and the conservation of electrical charges to find a better understanding of the SPT phases.  Note that even if the on-site symmetry of the Hamiltonian is non-Abelian, we can choose $G$ to be an Abelian subgroup of this symmetry, and apply the arguments for the charges corresponding to this Abelian subgroup.

\subsubsection{Example: $\mathbb{Z}_2\times \mathbb{Z}_2$ symmetry}\label{Sec:Z2}
In this section we consider the example of the group $\mathbb{Z}_2\times \mathbb{Z}_2$.  Formally, this group can be thought as the group of strings of two bits $b_ob_e$ with bitwise XOR as the group operation, that is the set  $\{00,01,10,11\}$.  This groups protects the Haldane phase of spin chain with spin-one systems, which includes states such the AKLT state \cite{AKLT}. In this case the on-site representation of this group is the set of four operators $\{I, e^{iS_x \pi}, e^{iS_y \pi}, e^{iS_z \pi} \}$, where $I$ is the identity operator, and $S_x$, $S_y$ and $S_z$ are spin matrices in three orthogonal directions. Note that for integer spins the $\pi$ rotations around three orthogonal axes commute with each other.   

As another example, consider the representation of this group on a spin chain with spin-half systems (qubits). We can group pairs of neighbor qubits, and consider each pair as one site.  Then, on each site formed from two qubits, the group  $\mathbb{Z}_2\times \mathbb{Z}_2$ can be represented by  
\beq
u(b_ob_e)=X^{b_o}\otimes X^{b_e} ,\ \  b_{e,o}\in\{0,1\}\ ,
\eeq
where $X$ denotes the Pauli $\sigma_x$ operator.  In other words, on each pair of qubits the group is  represented by one of the following operators
$$\{I\otimes I, X\otimes I, I \otimes X, X \otimes X \}\ .$$  

Let $C$ be a connected region with even number of qubits $2l$. Then, the action of group in this region is represented by  
\begin{align}
U^{(C)}(b_ob_e)& = X_\text{odd}^{b_o} X_\text{even}^{b_e}\ ,
\end{align}
where $X_\text{odd}=(X\otimes I)^{\otimes l}$ (or  $X_\text{even}=(I\otimes X)^{\otimes l}$) are the Pauli $\sigma_x$ operators acting on all odd (or even) qubits in region $C$.

This Abelian group has four elements, and hence has four different unitary 1-dimensional irreps, or equivalently four different charges. These four different charges can be   labeled by two bits $r_o,r_e\in\{0,1\}$. In particular, for the charge labeled by $r_or_e$ the group element $b_ob_e$ is represented by  $(-1)^{r_ob_o+r_eb_e}$.  
Then, the projector to the subspace with this charge in region $C$ is 
\bes
\begin{align}
\Pi_{r_or_e}^{(C)}&=\frac{1}{4}\sum_{b_o,b_e=0}^1 (-1)^{b_or_o+b_er_e} X_\text{odd}^{b_o} X^{b_e}_\text{even}  \\ &=(\frac{I+(-1)^{r_o}  X_\text{odd}}{2})(\frac{I+(-1)^{r_e}  X_\text{even}}{2})\ .
\end{align}
\ees
Note that $\frac{I+(-1)^{r_o}  X_\text{odd}}{2}$ is the projector to the subspace for which the total parity of  $X$ measurements on all odd qubits in region $C$ is $r_o$.   Similarly, $\frac{I+(-1)^{r_e}  X_\text{even}}{2}$ is the projector to the subspace for which the total parity of  $X$ measurements on all even qubits in region $C$ is $r_e$.  

We come back to this example in Sec.(\ref{Sec:Example}), where we discuss SPT phase of cluster state.

\subsection{Symmetric low-depth circuits}\label{Sec:low-depth}

Different SPT phases can be classified based on the equivalence classes of states induced by  the \emph{symmetric low-depth quantum circuits} \cite{Wen-Local}. According to this classification,  two states are in the same SPT phase if one can be approximately transformed to the other by a low-depth circuit $V=\prod_{k=1}^{l} V_{i}$ where each $V_{i}$ is a product of a set of unitaries which (i) act locally on non-overlapping regions of the system, and (ii) are invariant under the symmetry (That is satisfy $[V_i, U(g)]=0$, for all $g\in G$).  The circuit should be \emph{low-depth} in the sense that the {depth} $l$ times the maximum range of each unitary in the circuit is bounded by some range $R$, which in the thermodynamics limit is negligible compared to the system size. 

\begin{figure} [h]
\begin{center}
\includegraphics[scale=1.0]{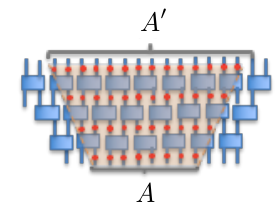}
\caption{The light-cone corresponding to a low-depth circuit. After applying the unitary, information which is initially localized in region $A$, will stay inside region $A'$, which is only slightly larger than $A$. }\label{fig:low-depth}
\end{center}
\end{figure}

An important feature of the low-depth circuits   is that  they have a \emph{light cone} (See Fig. \ref{fig:low-depth}).  That is if the circuit $V$ has range $R$, then for any  local operator $O_a$ which has support only on a single site $a$, the operator $V^\dag O_a V$ has support only on sites whose distance from site $a$ is, at most, $R$. 



\subsection{Measures of Entanglement and LOCC}\label{Sec:Per:Ent}

In recent years entanglement theory in general-and measures of entanglement in particular-  has found lots of applications in many-body systems (See e.g. \cite{Amico2008, Localizable, All1,All2, Paolo}). Many authors have studied how different measures of entanglement, such as negativity \cite{Vidal-Werner} or concurrence \cite{Wootters}, can detect certain properties of a many-body system. In this paper, however,  we are not concerned with particular measures of entanglement, because our results on SPT-Entanglement hold for all such functions.  Indeed, it turns out that to understand SPT phases, and in particular, the notion of SPT-Entanglement, it is useful to go beyond particular measures of entanglement, and take a more resource-theoretic  perspective to entanglement. 

In quantum information theory entanglement is defined as the property of states that cannot be generated with Local Operations and Classical Communications (LOCC).  These are basically all and the only transformations that can be implemented on a bipartite system shared between two distant local parties,  Alice and Bob, if the only interactions between them is through a classical channel.  These transformations include (i) local operations on each side, such as local unitary transformations, measurements, and discarding subsystems (partial trace) and (ii) classical communication between the two parties. Under this restricted class of operations, an untangled state cannot be transformed to an entangled one. Also, a given entangled state cannot be transformed to another arbitrary entangled state, if the latter state has more entanglement.  

Recall that a general bipartite mixed state $\rho^{AB}$ is unentangled, or \emph{separable}, if it can be written as $\rho^{AB}=\sum_i p_i \rho^A_{i}\otimes \rho^B_{i}$, that is a  convex combination of uncorrelated states $\rho^A_{i}\otimes \rho^B_{i}$, where $\{p_i\}$ is a probability distribution. Note that  using LOCC transformations it is possible to transform an uncorrelated state $\rho^A\otimes \rho^B$, to a correlated one $\sum_i p_i \rho^A_{i}\otimes \rho^B_{i}$. In other words, using LOCC a state with zero mutual information can be transformed to a state with non-zero mutual information. 

Unlike mutual information which can increase under classical communication, measures of entanglement can distinguish between  classical correlations and quantum correlations, i.e. those that cannot be generated via classical communication. In particular,  if using  LOCC a bipartite state $\rho^{AB}$ can be transformed to state $\sigma^{AB}$, then any measure of entanglement $E$ should be non-increasing in this transformation, i.e.  
\beq\label{def-meas}
\rho^{AB} \xrightarrow{\text{LOCC}}\sigma^{AB}\ \ \Longrightarrow  \ \   E(\rho^{AB})\ge E(\sigma^{AB}) .   
\eeq
Here, the  arrow $\xrightarrow{\text{LOCC}}$ means that the transformation from the first state to the second is possible via Local Operations and Classical Communication. 

In this paper, rather than focusing on a particular measure of entanglement,  we phrase our results in terms of interconvertability of states under LOCC. Note that  monotonicity of measures of entanglement under LOCC, i.e. Eq.(\ref{def-meas}), immediately implies that  
\begin{proposition}\label{prop-ent}
Suppose  a pair of states $\rho^{AB}$ and $\sigma^{AB}$ can be reversibly transformed to each other using LOCC, that is $\rho^{AB} \xrightarrow{\text{LOCC}}\sigma^{AB}$ and  $\sigma^{AB} \xrightarrow{\text{LOCC}}\rho^{AB}$. Then, any measure of entanglement $E$ takes the same value on these two states, i.e. $E(\rho^{AB})=E(\sigma^{AB})$ . 
\end{proposition}

In addition to the monotonicity under LOCC, it is usually assumed that a measure of entanglement vanishes on all unentangled (separable) states. In particular, for all product states $\rho^A\otimes \rho^B$, it should hold that
\beq\label{ref-zero}
E(\rho^{A}\otimes\rho^B)=0\ .   
\eeq
Note that because any unentangled state can be transformed to any other unentangled state via LOCC, proposition \ref{prop-ent} implies that all measures of entanglement should take the same value on all unentangled states. Therefore, Eq.(\ref{ref-zero}) is just a convention that fixes this value to be zero.

A well-known example of measures of entanglement is  \emph{negativity} \cite{Vidal-Werner} defined by
\beq
\mathcal{N}(\rho^{AB})\equiv \frac{\|{\rho^{AB}}^{\textbf{T}_{A}}\|_{1}-1}{2}\ ,
\eeq
where ${\rho^{AB}}^{\textbf{T}_{A}}$ is the operator obtained by partial transpose on system $A$ (relative to an arbitrary basis), and $\|\cdot \|_1$
is the sum of the absolute values of the eigenvalues of the operator. One can easily check that for a maximally entangled state of a pair of $d$-dimensional systems, i.e. state $\frac{1}{\sqrt{d}}\sum_{i=1}^d|ii\rangle$ the negativity is equal to $(d-1)/2$.

 Note that  the \emph{entanglement entropy} of pure bipartite states, defined as  the entropy of the reduced state of one side,  is non-increasing under LOCC in pure state to pure state transformations. However,  it can increase in the transformations between mixed states, and therefore it is not a valid measure of entanglement for mixed states. For instance, for the totally mixed state of a bi-partite system, which is clearly unentangled,   the entropy of the reduced states could be arbitrarily  large, depending on the dimension of the Hilbert space.   

\subsubsection{Example: Conditionally rotated states}\label{Sec:exa1}
Consider a bi-partite pure state $|\psi\rangle^{AB}$ shared between two distant parties, Alice and Bob. For any local unitary $U$ acting on the Bob' system, the two states $(I\otimes U)|\psi\rangle^{AB}$ and $|\psi\rangle^{AB}$, have the same amount of entanglement relative to any measure of entanglement $E$, that is $E((I\otimes U)|\psi\rangle^{AB})=E(|\psi\rangle^{AB})$. 

Now suppose   $\{U_k\}$ is an arbitrary set of unitary transformations acting on system $B$, and $\{p_k\}$ is a probability distribution over this set. Consider  the convex combination 
\beq
\Omega^{AB}=\sum_k p_k\   (I\otimes U_k)|\psi\rangle\langle\psi|^{AB}(I\otimes U^\dag_k)\ ,
\eeq
that is a mixture of rotated versions of state $|\psi\rangle^{AB}$, weighted by the probability distribution $\{p_k\}$. 
 
 Then, in general, for any measure of entanglement $E$ it holds that
\beq
E\Big(\Omega^{AB} \Big)\le E\Big(|\psi\rangle\langle\psi|^{AB} \Big) \ .
\eeq
This is because state $|\psi\rangle^{AB}$ can be transformed to state  $\Omega^{AB}$ via local operations, namely by applying unitary $U_k$ on system $B$ with probability $p_k$. In general, the left-hand side is strictly smaller than the right-hand side. In particular, we can always choose unitaries $\{U_k\}_k$ and a probability distribution $p_k$ such that state $\Omega^{AB}$ becomes unentangled, for all initial states $|\psi\rangle^{AB}$.

Next, consider state 
\beq
\Omega^{ABK}=\sum_k p_k\ |k\rangle\langle k|^{K}\otimes (I\otimes U_k)|\psi\rangle\langle\psi|^{AB}(I\otimes U^\dag_k) \ .
\eeq
Here, states  $\{|k\rangle\}$  are orthonormal states of a \emph{classical register} $K$, which keeps information about the unitary $U_k$ which is applied to system $B$.  Now suppose we give the systems $K$, and $B$ to Bob, and system $A$ to Alice. In other words, we partition these three systems as $A|KB$. Then, it can be easily seen that the bi-partite entanglement of state $\Omega^{ABK}$ relative to this partition is still equal to the entanglement of state $|\psi\rangle^{AB}$. In other words, both transformations $ |\psi\rangle^{AB} \xrightarrow{\text{LOCC}}\Omega^{ABK} $ and $\Omega^{ABK}  \xrightarrow{\text{LOCC}} |\psi\rangle^{AB} $ can be implemented via LOCC. In particular, to perform the latter transformation,  Bob can measure the value of register $K$, and if he finds the register in state $k$, then he applies unitary $U_k^\dag$ on system $B$. We conclude that for any measure of bi-partite entanglement $E$ for the partition  $A|KB$, it holds that
\beq\label{eqabove}
E(\Omega^{ABK})=E(|\psi\rangle^{AB})\ .
\eeq
One can easily check this formula for negativity $\mathcal{N}$, and show that it takes the same value on both states.

Next, suppose instead of partition $A|KB$, we consider the partition $AK|B$. That is we give the classical register $K$ to Alice, instead of Bob. What is the entanglement of state $\Omega^{ABK}$ relative to this partition? It can be easily seen that the entanglement remains unchanged. The reason is that  register $K$ has only classical information, and therefore Alice and Bob can transfer this information from one side to the other via classical communication. 

Note that in the above argument about the equivalence of bi-partite entanglement relative to $A|KB$ and $AK|B$ partitions, the monotonicity of entanglement measures under classical communication plays an important role. If instead of measures of entanglement, we consider a general measure of correlation, such as mutual information,  which can increase under classical communication, then the bi-partite correlation in state $\Omega^{ABK}$  will not be generally the same relative to these two different partitions.

\section{SPT-entanglement}\label{SPT-entanglement}

\subsection{Definition}

\begin{figure} [h]
\begin{center}
\includegraphics[scale=.33]{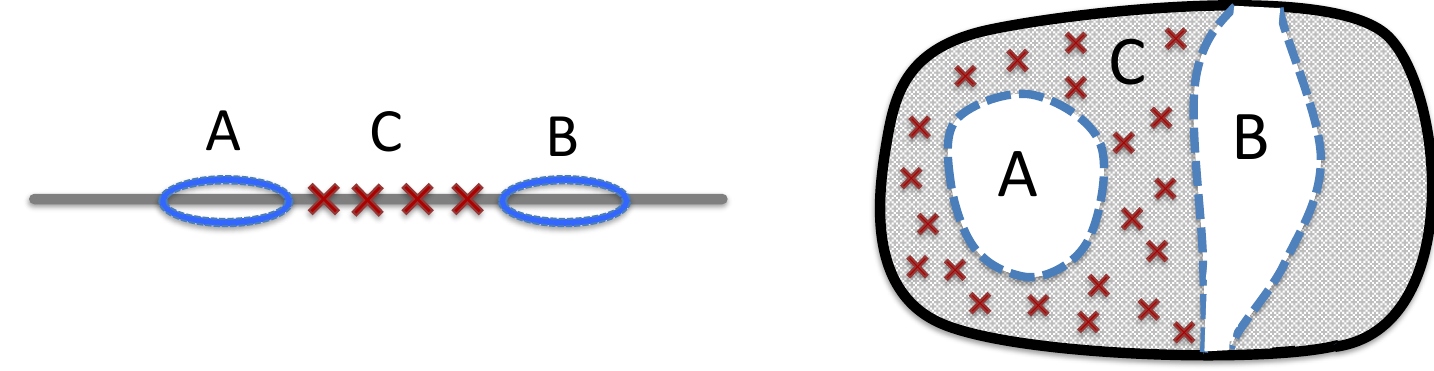}
\caption{Region $C$ is surrounded by regions $A$ and $B$ and the boundaries of the system. }\label{fig:Regions}
\end{center}
\end{figure}

Let $A$ and $B$ be two non-overlapping regions of the system, and $C$ be a connected region surrounded by $A$,  $B$ and the boundaries of the system (see Fig \ref{fig:Regions}).  Let $G$ be an Abelian symmetry which has on-site linear unitary representations on the sites of this system. Suppose we measure the total charge associated with the  symmetry $G$ in region $C$ and obtain charge $\kappa$. Here, each $\kappa$ corresponds to a distinct irreducible representation of $G$ as $g\rightarrow e^{i\kappa(g)}$. Let $\Pi_{\kappa}^{(C)}$ be the projector to the subspace of states with charge $\kappa$ in  region $C$ (See Eq.(\ref{aero})). If the system is in state $\rho$, then by measuring the total charge in  $C$  the charge $\kappa$ is obtained with the probability $p_{\kappa}=  \text{tr}(\rho \Pi^{(C)}_{\kappa})$, and given this outcome, the reduced state of $AB$ after the charge measurement will be  $\rho_{{\kappa}}^{(AB)}\equiv\text{tr}_{\overline{AB}}(\Pi^{(C)}_{\kappa} \rho)/{p_{\kappa}}$, where the trace is over all the sites in the system except those in $A$ and $B$.

Consider the state
 \bes\label{state}
 \begin{align}
\Omega^{(AB|C)}({\rho})&\equiv \sum_{\kappa} p_{\kappa} |\kappa\rangle\langle\kappa|^{(K_{C})}\otimes \rho_{\kappa}^{(AB)} \\ &=\sum_{\kappa}   |\kappa\rangle\langle\kappa|^{(K_{C})}\otimes \text{tr}_{\overline{AB}}(\Pi^{(C)}_{\kappa} \rho)\ ,
\end{align}
\ees
where $\{|\kappa\rangle\}$ are the orthonormal states of a classical register $K_{C}$ which keeps the information about the outcome of charge measurement  in region $C$.  State $\Omega^{(AB|C)}({\rho})$ describes the average joint state of regions  $A$ and $B$ and the register $K_{C}$ after the charge measurement.

Note that the super-operator
\begin{equation}
\Omega^{(AB|C)}(\cdot)\equiv\sum_{\kappa} |\kappa\rangle\langle\kappa|^{(K_{C})}\otimes \text{tr}_{\overline{AB}}(\Pi^{(C)}_{\kappa} \cdot)
\end{equation}
is a trace-preserving completely positive map, i.e. a quantum channel. This is basically the map which traces over all the degrees of freedom in the system except (i) the total charge in region $C$, and (ii) local degrees of freedom in regions  $A$ and $B$. Therefore, an alternative interpretation of state $\Omega^{(AB|C)}({\rho})$ is that it is  the reduced state  of the system relative to the algebra generated by the local observables in regions $A$ and $B$ and the observable corresponding to the total charge in region $C$. 

\begin{figure} [h]
\begin{center}
\includegraphics[scale=.30]{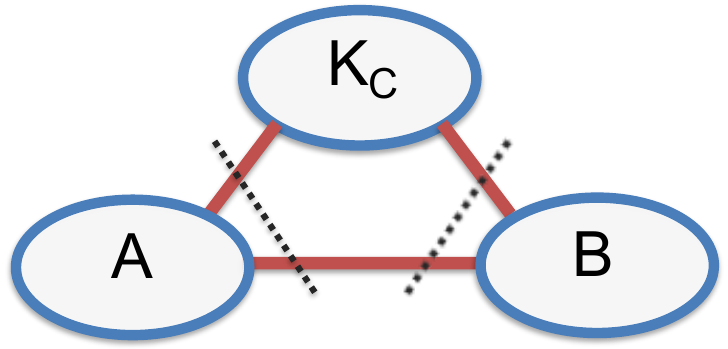}
\caption{State $\Omega^{(AB|C)}({\rho})$ describes the joint state of the  local degrees of freedom at regions $A$ and $B$, and register $K_{C}$, which corresponds to the total charge in region $C$. The SPT-Entanglement is defined as the bipartite entanglement of this state relative to the partitions $A|K_C B$ or $A K_C|B$. }\label{fig:Regions2}
\end{center}
\end{figure}

Now we are ready to define the notion of SPT-Entanglement:  

\begin{definition}
Let $A$ and $B$ be two non-overlapping regions of the system,  and $C$ be a connected region surrounded by $A$ and $B$ and the boundaries of the system (see Fig \ref{fig:Regions}). The SPT-Entanglement  of state $\rho$ between $A$ and $B$ relative to $C$ is defined as the bipartite entanglement of state $\Omega^{(AB|C)}({\rho})$ relative to $AK_{C}|B$ partition, or equivalently, relative to $A|K_{C}B$ partition (See Fig \ref{fig:Regions2}). 
\end{definition}
Several important remarks are in order: First, since $K_{C}$ is a classical register, it can be transferred from one local party to the other via classical communication. Therefore,  the bipartite entanglement of state $\Omega^{(AB|C)}({\rho})$ relative to the partitions $AK_{C}|B$ and $A|K_{C}B$ is the same. Second,  we are interested in the SPT-Entanglement in the limit where regions $A$ and $B$ are large and far from each other. Third, the Abelian group $G$ could be a subgroup of a (possibly non-Abelian) group that protects the phase. 



Note that in this definition we talk about bipartite entanglement of state $\Omega^{(AB|C)}({\rho})$, without specifying a particular measure of entanglement. This is because the results presented in this paper hold for all measures of entanglement, and we do not need to specify one. But, to have an explicit example, we can quantify SPT-Entanglement using negativity, which yields the function 
\begin{align}
N^{(AB|C)}(\rho)&\equiv \mathcal{N}(\Omega^{(AB|C)}(\rho))\nonumber\\ &=\frac{1}{2}\left[\sum_{\kappa}\left\| \left[\text{tr}_{\overline{AB}}(\Pi^{(C)}_{\kappa} \rho)\right]^{\textbf{T}_{A}}   \right\|_{1}-1\right],
\end{align}
where $\textbf{T}_{A}$ denotes the partial transpose,  and $\|\cdot\|_{1}$ is the trace norm, i.e. the sum of the absolute value of eigenvalues. As we will see in Sec.(\ref{Sec:strin}), this function can be expressed in terms of the string order parameters.




\subsection{Main Properties: Summary of results}\label{Sec:main}

In the following we present a summary of the main properties of the SPT-Entanglement.

\indent\textbf{(i)} It is clear from the definition of SPT-Entanglement that applying a symmetric unitary which acts non-trivially only on one of the three regions $A$, $B$ and $C$, does not change the SPT-Entanglement. On the other hand, a symmetric unitary with support in more than one region, can change the SPT-Entanglement.  However, we prove that if the symmetric unitary is a low-depth circuit then the SPT-Entanglement cannot change drastically. More precisely, we show that the effect of a low-depth symmetric circuit on the SPT-Entanglement can be bounded based on the effect of small deformations of the boundaries of regions $A$ and $B$ on this quantity: If by deforming the boundaries of these regions the SPT-Entanglement cannot change considerably, then it also cannot change considerably under low-depth symmetric circuits (See theorem \ref{Thm:LOCC-transmain}). Note that SPT-Entanglement can change drastically under low-depth \emph{non-symmetric} unitaries. In the following, we explain the intuition behind this result. See Sec.(\ref{Sec:effect}) for a more formal (and slightly different) argument.

\begin{figure} [h]
\begin{center}
\includegraphics[scale=.45]{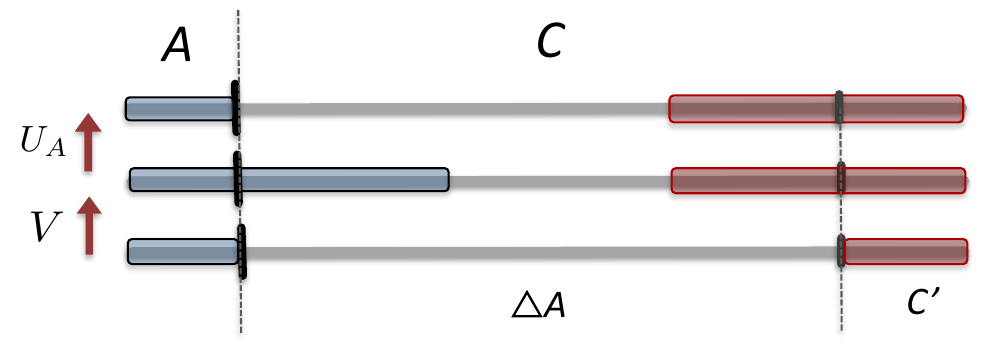}
\caption{
Under the symmetric low-depth circuit $V$ quantum information and charge from  region $A$ can leak to  region $\Delta A$, that is a subset of sites in region $C$ which are close to the sites in region $A$.  
Applying the symmetric unitary  $U_{A}$ on the region $A'=A \cup \Delta A$, Alice can restore the leaked charge and quantum information from region  $A$ back to this region.  }\label{fig:core}
\end{center}
\end{figure}

Suppose a symmetric low-depth circuit $V$ is applied on a system in the initial state $\rho$, and transforms it to state  $V\rho V^\dag$. Under this transformation quantum information and charge from  region $A$ can leak to  region $\Delta A$, which is a subset of sites in region $C$ which are close to the sites in region $A$  (See Fig.(\ref{fig:core})).  Similarly, the charge and quantum information in region $B$ can leak to $\Delta B$, a subset of sites in region $C$ which are close to the sites in region $B$. Let $C'=C\setminus ( \Delta A\cup  \Delta B)$ be the set of sites in region $C$ which are outside $\Delta A$ and $\Delta B$, and therefore are not close to $A$ or $B$. Then, under the effect of circuit $V$ the charge from region $C'$ can also leak to regions $\Delta A$ and $\Delta B$  (See Fig.(\ref{fig:core})).

Now suppose Alice is given all the sites in region $A'=A\cup \Delta A$, and Bob is given all the sites in region $B'=B\cup \Delta B$. Furthermore, suppose they know the charge in region $C'$ (for state $V\rho V^\dag$). In other words, assume they are given state $\Omega^{(A'B'|C')}(V\rho V^\dag)$. Then, by applying a symmetric unitary  on  region $A'$, Alice can undo the effect of $V$ on $A$ and restore the leaked charge and quantum information back to this region. Similarly, by applying a local symmetric unitary on $B'$, Bob can restore the charge and quantum information leaked from region $B$ back to this region. Next, by measuring the charges in regions $\Delta A$ and $\Delta B$, and adding these charges to the charge in region $C'$, whose value is given to them, Alice and Bob can find the charge in region $C$ for the original state $\rho$, and construct the state $\Omega^{(AB|C)}(\rho)$. This proves that there exists a LOCC protocol which transforms $\Omega^{(A'B'|C')}(V\rho V^\dag)$ to state $\Omega^{(AB|C)}(\rho)$, which, in turn, implies that for any measure of entanglement $E$, 
\beq\label{Eq.3.4}
E(\Omega^{(A'B'|C')}(V\rho V^\dag))\ge E(\Omega^{(AB|C)}(\rho))\ ,
\eeq
where the entanglement is defined relative to $A|K_CB$ partition or  $AK_C|B$ partition. 
 
Similarly, using the fact that if $V$ is a symmetric low-depth circuit,  then $V^\dag$ is also a symmetric low-depth circuit, we can find an upper bound on the SPT-Entanglement of state $V\rho V^\dag$ for regions $A'$ and $B'$, based on the  SPT-Entanglement  of the original state $\rho$ for regions which  are slightly larger than $A'$ and $B'$. 

\indent\textbf{(ii)} According to the classification of SPT phases, the trivial phase is the set of states which can be transformed to a product state $\rho=\bigotimes_i\rho_i$ using symmetric low-depth circuits  \cite{Wen-Local}.  As we saw above, the effect of low-depth symmetric circuits on the SPT-Entanglement can be bounded by the effect of small deformations of the boundaries of regions $A$ and $B$ on this quantity. But, clearly,  for a product state $\bigotimes_i \rho_i$ the SPT-entanglement between any non-overlapping regions $A'$ and $B'$ is zero. It follows that,  in the limit where two regions $A$ and $B$ are far from each other compared to the correlation length, the SPT-Entanglement is always zero in the trivial phase. Applying this general result to the negativity,  as an example of  measures of entanglement, we find
\begin{equation} \label{order}
\text{Trivial phase}\  \Longrightarrow\ \lim\ \sum_{\kappa}\left\| \left[\text{tr}_{\overline{AB}}(\Pi^{(C)}_{\kappa} \rho)\right]^{\textbf{T}_{A}}   \right\|_{1}-1=0\ ,
\end{equation}
where $\textbf{T}_{A}$ denotes the partial transpose,  and $\|\cdot\|_{1}$ is the trace norm, and we are looking to the limit where $A$ and $B$ are far from each other.

Therefore, a non-vanishing SPT-Entanglement between two distant regions, in  the limit where the distance between them is large, indicates the presence of SPT order. We show that this method can always detect all non-trivially SPT ordered phases in 1-dimensional spin systems.  Another way to interpret this result is that it  gives  a criterion that should be satisfied by any state which can be prepared by symmetric low-depth  circuits. Note that,  this method works even in the absence of translational symmetry (See \cite{Chen2} for a different approach). 

\indent\textbf{(iii)}  The above result implies that the SPT-Entanglement remains constant throughout the trivial phase. Making an assumption about the decay of correlations in the system, we can extend this result to all SPT phases, and show that in the limit where regions $A$ and $B$ are large and far from each other compared to the correlation length, SPT-Entanglement is a \emph{universal} quantity in all dimensions, i.e. it remains constant throughout the phase  (See corollary \ref{Thm1}). Similar to the argument that shows topological entanglement entropy remains constant throughout a phase \cite{Kitaev,Lev-Wen}, our argument is based on a reasonable, but unproven  assumption: Since SPT-Entanglement  increases monotonically with the sizes of regions $A$ and $B$ (See  proposition \ref{momont:Thm}), given that  the correlations are short range and the system is sufficiently homogenous (e.g., translational invariant),  it seems reasonable to assume that in the limit where regions $A$ and $B$ are large, SPT-Entanglement is saturated, i.e. it does not increase if we make $A$ and $B$ a bit larger. Making this assumption, and using Eq.(\ref{Eq.3.4}) which bounds the effect of low-depth symmetric circuits on SPT Entanglement,  we find that SPT Entanglement remains invariant under symmetric low-depth circuits, and hence is constant throughout a SPT phase.

 Note that in the case of 1-dimensional systems explicit calculation of SPT-Entanglement in the MPS framework confirms this assumption. However, the validity of this assumption in the higher dimensions  is not clear, and remains an open problem.


\indent\textbf{(iv)}  We directly calculate  SPT-Entanglement of 1-dimensional systems in the Matrix Product State (MPS) framework, and  show that  in the limit where  $A, B$ and $C$ are  large compared to the correlation length,  SPT-Entanglement  is:  (1) independent of the sizes of these regions, (2) constant throughout all SPT phases, and (3) zero if and only if the system is in the trivial phase. Furthermore, we show that the value of SPT-Entanglement in the SPT ordered phases is related to a physical property of the SPT phase, namely the edge mode degeneracy associated with each edge of an open chain.  More precisely, we show that in the limit  where $A$ and $B$ are large and far from each other the SPT-Entanglement is equal to the entanglement of a  maximally entangled state of a pair of $d_{[\omega]}$-dimensional systems, where $d_{[\omega]}$ is the dimension of the projective irreducible representations of the  group  $G$ in the cohomology class ${[\omega]}$ which  characterizes the SPT phase of the system  \cite{Clas-Wen,Clas-Wen2,Pollman-Turner2, Schuch-Cirac}. Note that, in general,  irreducible representations which belong to the same cohomology class  have different dimensions. However, interestingly,  we find that  for  \emph{Abelian} groups all such irreducible representations  have the same dimensions (See lemma \ref{lem-new-dim}).  It turns outs that the parameter $d_{[\omega]}$ has a natural interpretation as the edge mode degeneracy associated with each edge of an open chain \cite{Else2012,prep} (Note that this parameter is only defined for 1-dimensional systems). 

 According to the classifications of the SPT phases of 1-dimensional systems,  the equivalence class $[\omega]$ uniquely determines the SPT phase  \cite{Clas-Wen,Clas-Wen2,Pollman-Turner2, Schuch-Cirac}.   Therefore, in the case of 1-dimensional systems direct calculation of SPT-Entanglement confirms our general result that SPT-Entanglement is constant throughout a SPT phase. 

It follows from this result that, using different measure of entanglement to quantify SPT-Entanglement we can construct different order parameters to detect SPT order of 1-dimensional systems.  In particular, suppose  we quantify the SPT-Entanglement using negativity. Then, since the negativity of a maximally entangled state of a pair of $d_{[\omega]}$-dimensional systems is $(d_{[\omega]}-1)/2$,  this result implies that  in the limit where  $A,B$ and $C$ are much larger than the correlation length, $\lim N^{(AB|C)}(\rho)=(d_{[\omega]}-1)/2$, or equivalently
\begin{equation} \label{order}
\lim\ \sum_{\kappa}\left\| \left[\text{tr}_{\overline{AB}}(\Pi^{(C)}_{\kappa} \rho)\right]^{\textbf{T}_{A}}   \right\|_{1} -1= d_{[\omega]}-1.
\end{equation}
A 1-dimensional system is in a non-trivial phase  iff the corresponding cohomology class $[\omega]$ is non-trivial  \cite{Clas-Wen,Clas-Wen2,Pollman-Turner2, Schuch-Cirac}, or equivalently iff $d_{[\omega]}>1$. Therefore, the left-hand side of Eq.(\ref{order}) can serve as an order parameter whose value for the large blocks of $A$, $B$, and $C$ not only detects the presence of SPT order, but also reveals the dimension of the irreducible representations in the equivalence class $[\omega]$ which characterizes the phase. For example, in the case of Haldane phase where $d_{[\omega]}=2$, the left-hand side of Eq.(\ref{order}) converges to 1, while it converges to zero in the trivial phase. In Sec.(\ref{Sec:Example}) we check this result in the case of the  cluster Hamiltonian. 


\subsection{Connection with the string order parameters}\label{Sec:strin}
Remarkably, it turns out that the notion of  SPT-Entanglement is closely related to the concept of string order parameters, which have been the traditional tool for detecting the SPT order \cite{str-order, Hidden-Sym}. 

The string order parameters are the expectation values of state for string operators, in the following form 
\bes
 \begin{align}
 s_{kl}(g)&\equiv \Tr(\rho\ X^{(A)}_{k}\bigotimes_{j\in C} u_j(g)\otimes Y^{(B)}_{l}) \ ,\\ &=\Tr(\rho\ X^{(A)}_{k}\otimes U^{(C)}(g)\otimes Y^{(B)}_{l})
 \end{align}
 \ees
  where  $\{X^{(A)}_{k}\}_k$ and $\{Y^{(B)}_{l}\}_l$ are  basis  for the space of local operators acting on $A$ and $B$ respectively, and $g\rightarrow u_j(g)$ is the  representation of the group element  $g\in G$ on site $j$. 
The relation between the string order parameters and  the SPT-Entanglement can be established using the notion of Fourier transform for Abelian groups. Let 
\beq
\tilde{s}_{kl}(\kappa)=\frac{1}{|G|}\sum_{g\in G} e^{-i\kappa(g)}\ s_{kl}(g)\ ,
\eeq
be the Fourier transform of the string order parameter $s_{kl}(g)$. Then, using $\Pi_{\kappa}^{(C)}=|G|^{-1}\sum_{g\in G} e^{-i\kappa(g)} U^{(C)}(g)$, and the definition of state $\Omega^{(AB|C)}(\rho)$ in Eq.(\ref{state}), we can easily see
\begin{align}\label{Fourier}
\tilde{s}_{kl}(\kappa)= \Tr\left(\left[\ |\kappa\rangle\langle\kappa|^{(K_{C})}\otimes X^{(A)}_{k}\otimes Y^{(B)}_{l}\right] \Omega^{(AB|C)}(\rho)\right).
\end{align}
Using these equations for different charges $\kappa\in Q$, and different local operators $ X^{(A)}_{k}$ and $Y^{(B)}_{l} $, one can reconstruct  state $\Omega^{(AB|C)}({\rho})$, and hence find the SPT-Entanglement of  $\rho$, between $A$ and $B$ relative to $C$.

In particular, suppose the local operators $ \{X^{(A)}_{k}\}_k$ and $\{Y^{(B)}_{l} \}_l$ form an orthonormal basis for the space of local operators acting on $A$ or $B$, respectively, such that $ \Tr(X^{(A)}_{k}  {X^{(A)}_{k'}}^\dag)=\delta_{k,k'}$ and $ \Tr(Y^{(B)}_{l}  {Y^{(B)}_{l'}}^\dag)=\delta_{l,l'}$. Then, from Eq.(\ref{Fourier}) we find
\begin{align}
\Omega^{(AB|C)}(\rho)=\sum_\kappa \sum_{k,l} \tilde{s}_{k,l}(\kappa)\ |\kappa\rangle\langle\kappa|^{(K_{C})}\otimes {X_{k}^{(A)}}^\dag\otimes {Y_{l}^{(B)}}^\dag .
\end{align}
Therefore, choosing any measure of entanglement, we can quantify the SPT-Entanglement of state in terms of the string order parameters. In particular, if we use the negativity of entanglement, we find 
\beq
N^{(AB|C)}(\rho)=\frac{1}{2}[\sum_{\kappa}\Big \| \sum_{k,l}  \tilde{s}^\ast_{k,l}(\kappa)\ [{X_{k}^{(A)}}\otimes {Y_{l}^{(B)}}^T] \Big\|_{1}-1]\ ,
\eeq
where $T$ is the transpose relative to an arbitrary basis. As we explained before, in the case of 1-dimensional systems, and in the limit where $A$, $B$ and $C$ are large this quantity converges to $(d_{[\omega]}-1)/2$, where $d_{[\omega]}$ is the dimension of the irreps in the cohomology class $[\omega]$ that characterizes the SPT phase. 

We conclude that, although  each single string order parameter $s_{kl}(g)$ alone does not have any information about the SPT phase of the system \cite{Cirac-String}, but  together they  provide  enough information to find the SPT-Entanglement, and hence detect the SPT order. Indeed, even further, they can be used to find the dimension $d_{[\omega]}$ corresponding to the phase of system. 

See \cite{Else2013} for another recent approach to string order parameters, which applies for Abelian groups, provided that the cohomology class describing the phase satisfies a condition called  \emph{maximal non-commutativity}.

\subsection{Remarks on the definition of SPT-Entanglement}
There are several assumptions in the definition of SPT-Entanglement, and one may wonder if these assumptions are crucial or not. Here, we discuss more about the importance of these assumptions.

\subsubsection{Why Abelian symmetries?}
In both arguments that we present in the paper, i.e. the 1-dimensional argument based on MPS in Sec.(\ref{Sec:1-d}), and the general argument based on symmetric  low-depth circuits in Sec.(\ref{Sec:D}), the fact that the symmetry under consideration is Abelian is crucial. It is interesting to see how in these  arguments, which are completely different from each other, this assumption is needed in different ways. For instance, in the argument in terms of symmetric  low-depth circuits, we use the fact that the charges corresponding to an Abelian group are additive. That is the total charge in one region can be written as the sum of the charges in its subregions.  On the other hand, in the case of 1-dimensional systems, where the SPT phases can be classified in terms of the cohomology class of projective representations of the symmetry, to prove our result we use another interesting fact about Abelian symmetries: All irreducible projective representations of an Abelian group in the same cohomology class have the same dimension (See lemma \ref{lem-new-dim}). 

Of course the notion of charge measurement can be extended to the case of non-Abelian groups, where the charge can be interpreted as the label for different irreps of the group. However, it can be easily shown that for non-Abelian groups, if we measure the charge in region $C$, then regions $A$ and $B$ can become entangled, even if the system is in the trivial phase (Note that unlike the case of Abelian charges, to measure a non-Abelian charge in region $C$ we need to act on all the sites in this region collectively, and this can create entanglement between $A$ and $B$).   It follows that  for non-Abelian charges the SPT-entanglement  is not a  universal quantity. 

We remind the reader that the Abelian group in the definition of SPT-Entanglement could be a subgroup of a possibly non-Abelian group that protects the phase.

\subsubsection{Entanglement versus correlation}
Whenever one uses entanglement measures to study a phenomenon in the context of many-body systems, one can ask if entanglement is really the relevant property to consider.  In other words, why we should look at a measure of entanglement and not other measures of correlations such as mutual information, which are often easier to calculate (Note that in the special case of bipartite pure states, mutual information is just twice the entanglement entropy, and therefore is a measure of entanglement. In other words, in this special case entanglement and correlation can be thought to be the same properties). 

As we discussed in Sec.(\ref{Sec:Per:Ent}) the defining property of measures of entanglement which distinguishes them from measures of correlations, is their monotonicity under classical communication. But, in the context of many-body systems  it is not often clear that why this distinction between classical and quantum communication is relevant at all.

Interestingly, the monotonicity of measures of entanglement under classical communication plays a clear role in both arguments that we present to prove the properties of SPT-Entanglement (i.e. the general argument based on low-depth circuit in Sec.(\ref{Sec:D}), and the 1-dimensional argument in Sec.(\ref{Sec:1-d})).  One of the reasons  that monotonicity of entanglement measures under classical communication is important in these arguments is the following fact: When we measure the charge in region $C$ then regions $A$ and $B$ can become correlated. In particular, if the total charge in $A$, $B$ and $C$ is known, then  measuring the charge in region $C$ can induce correlations between the charges in region $A$ and region $B$ (They should sum up to a known value, and therefore they should be correlated). But this is a classical correlation between the charge degrees of freedom of regions $A$ and $B$. Therefore, using measures of entanglement we can automatically filter out this kind of classical correlations, which are induced by the measurement in region $C$.  

\subsubsection{SPT-Entanglement versus Localizable Entanglement}

For any given state of a many-body system the \emph{Localizable Entanglement} between any two sites in the system is defined as the maximal amount of entanglement that can be created (localized) between the two sites by performing local (single-site) measurements on all other sites in the system \cite{Localizable, Stephen2009}.  Unlike SPT Entanglement, Localizable Entanglement is determined solely by the state of system, and does not depend on the symmetry under consideration. As we will see in the following example, this means that Localizable Entanglement can be non-zero even in the trivial phase, and its value does not tell us anything about the SPT order of the system. 

As an example, consider two parallel spin chains which are both, for instance, in 1-d cluster state (See Sec. \ref{Sec:Example} for definition of cluster state). Suppose each \emph{site} is formed from four qubits, i.e. a pair of neighbor odd and even qubits from one chain, and a pair of neighbor odd and even qubits from the other chain.  It can be easily shown that the Localizable Entanglement for each cluster state is 1-ebit, i.e.,  equal to the entanglement of a maximally entangled state of a pair of qubits. It follows that the Localizable Entanglement for two copies of cluster state is 2-ebits. 

On the other hand, depending on how one defines the symmetry, this system may or may not be in a SPT-ordered phase: Each cluster state has a $\mathbb{Z}_2\times \mathbb{Z}_2$ symmetry (See Sec. \ref{Sec:Example}).  Now we can treat the two copies of cluster state independently, and interpret the total symmetry as $\mathbb{Z}_2\times \mathbb{Z}_2\times \mathbb{Z}_2 \times \mathbb{Z}_2$. Then, as a whole system, the two copies of cluster state are in a non-trivial SPT phase protected by this symmetry.  In this case one can easily show that the SPT-Entanglement of the system is 2-ebits. On the other hand, if we assume the same group element acts on both chains, we can interpret the symmetry of the two copies of cluster state as $\mathbb{Z}_2\times \mathbb{Z}_2$. Then, in this case the two copies of cluster state, as a whole system, are in the trivial SPT phase of $\mathbb{Z}_2\times \mathbb{Z}_2$. In other words, there is a low-depth circuit which respects this $\mathbb{Z}_2\times \mathbb{Z}_2$ symmetry, and transforms the two copies of cluster state to a product state. Crucially, it can be shown that, unlike Localizable Entanglement, SPT-Entanglement of two copies of cluster state vanishes for this symmetry, as it should for any state in the trivial SPT phase.  

This example shows that, in general, a nonzero Localizable Entanglement does not imply anything about the SPT order of the system. Therefore, Localizable Entanglement cannot be used as an order parameter to detect SPT-ordered phases. 

Finally, it is worth noting that even if the system is in a SPT-ordered phase, the Localizable Entanglement between any two distant sites can be zero. In other words, to see the entanglement corresponding to the SPT-ordered phase, one may need to look at large regions of the system, and so strictly speaking, entanglement is not \emph{localizable}.

\subsubsection{Topology of the three regions}

Another assumption which is made in the definition of SPT Entanglement is that region $C$ is surrounded by regions $A$, $B$ and the boundaries of the system. This condition guarantees that if we apply a local symmetric unitary to the system, then the charge in this region either goes to region $A$, or region $B$. Therefore, having access to these regions one can find the original charge in region $C$. Again, one can easily construct examples to show that if this condition does not hold, then the SPT-Entanglement will not be a universal quantity.

\section{Example: Perturbed Cluster Hamiltonian}\label{Sec:Example}

The cluster Hamiltonian for qubits (spin-half systems) on a ring is defined by
\begin{align}
H_\text{clus}&= - \sum_i Z_{i-1} X_i Z_{i+1} =- \sum_i K_i \ , 
\end{align}
where $X$ and $Z$ denote the Pauli $\sigma_x$ and $\sigma_z$ operators,  and $K_i=Z_{i-1} X_i Z_{i+1}$.  For simplicity we assume the ring has even number of qubits (The cluster Hamiltonian can also be defined on an open chain, where one can add two commuting boundary terms to remove the degeneracy (See e.g. \cite{Doherty2009})). It can be easily seen that all the terms $K_i$ commute with each other, and their eigenvalues are $\pm 1$. Hence  the Hamiltonian is gapped, and exactly solvable. Indeed, using the standard results in the stabilizer formalism, it can be easily shown that $H_\text{clus}$ has a unique ground state $|\Psi_\text{clus}\rangle$, called the (1-dimensional) \emph{cluster} state, which is the common eigenvector of all $K_i$ operators with eigenvalue $1$, that is 
\beq\label{cond231}
K_i |\Psi_\text{clus}\rangle=|\Psi_\text{clus}\rangle\ .
\eeq
This state plays an important role in the Measurement-Based Quantum Computation (MBQC) \cite{raussendorf2003measurement}.  The cluster  Hamiltonian and its perturbed versions  have been extensively studied in the recent years. In particular, for the case where the  Hamiltonian is perturbed by uniform magnetic field in $x$ direction, i.e. by the term $B\sum_i X_i$,  it has been shown that the Hamiltonian has  quantum phase transition at $|B|=1$ \cite{Doherty2009, Pachos2004three, Stephen2009}.

The cluster Hamiltonian commutes with the unitaries $\bigotimes_{i\in \text{even}} X_i$ and $\bigotimes_{i\in \text{odd}} X_i$, that is all Pauli $x$ operators acting on even qubits or odd qubits, respectively. These unitaries clearly  form a representation of the group $\mathbb{Z}_2\times\mathbb{Z}_2$. By grouping pairs of neighbor qubits together,  we can interpret each pair of odd and even neighbors as a \emph{site}.  Then, for any such pair of qubits, unitaries $X\otimes I$ and $I\otimes X$  generate an on-site representation of $\mathbb{Z}_2\times\mathbb{Z}_2$ \cite{Else2012,ElseBartlettDoherty} (See Sec.(\ref{Sec:Z2})). 

As we discussed before, according to the classification of SPT phases in 1-dimensional spin systems,  there is only one non-trivial SPT phase protected by $\mathbb{Z}_2\times\mathbb{Z}_2$ symmetry, and it has been shown that the cluster states is in this non-trivial phase \cite{Else2012,ElseBartlettDoherty}. (Note that in the absence of the symmetry, the cluster state can be transformed  to a product state via a low-depth circuit, and therefore it does not have intrinsic topological order). 
Remarkably, this property of cluster state has been linked to the fact that it is a computational resource in the context of MBQC, and more specifically, to the fact that it can be used as a \emph{quantum wire} \cite{Else2012,ElseBartlettDoherty, Stephen2009}. The computational power of cluster state is also closely related to the fact that it has a non-zero SPT-entanglement. 

We have seen that in the case of 1-dimensional systems and Abelian symmetries, the SPT-Entanglement of state is determined by $d_{[\omega]}$, the dimension of the irreducible  projective representations in the cohomology class $[\omega]$ corresponding to the SPT phase of  the system. In the case of group $\mathbb{Z}_2\times\mathbb{Z}_2$, the dimension of these irreducible representations are either one or two, corresponding to the trivial and non-trivial phase (The set of Pauli operators together with the identity operator forms an irreducible projective representation of $\mathbb{Z}_2\times\mathbb{Z}_2$ with dimension two). Therefore, for the non-trivial SPT phase protected by $\mathbb{Z}_2\times\mathbb{Z}_2$, we have $d_{[\omega]}=2$. 
Hence, according to our general results, the  SPT-Entanglement of 1-dimensional cluster state should be equal to the entanglement of a maximally entangled state of a pair of qubits. 

In the following, we examine this claim in more details. Indeed, we show that the claim holds true for the grounds states of a general class of Hamiltonians obtained by perturbing the cluster Hamiltonian.   Note that this result can be interpreted as an independent proof of the fact that the cluster state is in the non-trivial SPT phase protected by $\mathbb{Z}_2\times\mathbb{Z}_2$.

\subsection{SPT-Entanglement for cluster state}
To calculate SPT-Entanglement, first recall from Sec.(\ref{Sec:Z2}) that the group $\mathbb{Z}_2\times\mathbb{Z}_2$ has four possible charges, corresponding to the four one-dimensional irreducible representations of this group. These charges can be labeled by two bits $r_or_e\in\{00,01,10,11\}$, such that the group element $b_ob_e\in\{00,01,10,11\}$ is represented by $(-1)^{r_o b_o+r_e b_e}$. 

Let $C$ be a connected region with even number of qubits. Then, the projector to the subspace with charge labeled by $r_or_e$ in region $C$ is given by
\beq\label{defdef1}
\Pi^{(C)}_{r_or_e}=\big(\frac{I+(-1)^{r_o}X_\text{odd}^{(C)}}{2}\big) \big(\frac{I+(-1)^{r_e}X_\text{even}^{(C)}}{2}\big)\ ,
\eeq
where 
\beq
X_\text{even}^{(C)}=\prod_{\substack{i\in C\\
                 i: \text{even} }} X_i ,\ \ \  \text{and}\ \ \   \ X_\text{odd}^{(C)}=\prod_{\substack{i\in C\\
                  i: \text{odd} }}
\eeq
are the product of $X$ operators on even and odd qubits in region $C$, respectively.  In other words, measuring the total charge in region $C$ is equivalent to measuring both operators $X_\text{even}^{(C)}$ and $X_\text{odd}^{(C)}$. Therefore, the charge measurement can be done, for instance, by measuring $X$ operators on all qubits in region $C$, and then looking to the total parity of the outcomes of these measurements for odd qubits (which determines the bit $r_o$), and even qubits (which determines the bit $r_e$). Incidentally, this is how the identity gate is performed on the 1-dimensional cluster state in the MBQC (See Sec.(\ref{Sec:iden}) for further discussion). 

Suppose $C$ is surrounded by two regions $A$ and $B$ in the left side and the right side respectively, where each region contains at least two qubits (See Fig.(\ref{fig:Regions})). Let $|\Phi^{(AB)}_{r_or_e}\rangle$ be the post-measurement joint state of regions $A$ and  $B$ corresponding to the charge labeled by $r_or_e$ in region $C$. Then, for the case of cluster state $\Psi_\text{clus}$, the state $\Omega^{(AB|C)}(\Psi_\text{clus})$ defined in Eq.(\ref{state}) is given by
\begin{align}
\sum_{r_o,r_e=0}^1 p_{r_or_e} |r_or_e\rangle\langle r_or_e|^{K_C}\otimes \text{tr}_{\overline{AB}}(\Pi^{(C)}_{r_0 r_1} |\Psi_\text{clus}\rangle\langle\Psi_\text{clus}|)\ ,
\end{align}
where $p_{r_or_e}$ is the probability of outcome $r_or_e$ in the charge measurement.

Then, using the nice properties of the cluster state, one can easily show that (i) each of these four charges are obtained with equal probability $1/4$, and (ii) for each particular charge $r_or_e\in\{00,01,10,11\}$, region $A$ is entangled with  region $B$  and this entanglement is equal to the entanglement of a maximally entangled state of a pair of qubits (A general proof is presented below). More precisely, assuming  the outcome of the charge measurement in region $C$ is $r_or_e$, then up to local unitaries on $A$ and $B$, the joint state of these regions is given by
\beq\label{cluster-ent}
|\Phi^{(AB)}_{r_or_e}\rangle= (I_\textbf{a}\otimes X_\textbf{b}^{r_o}Z_\textbf{b}^{r_e}H_\textbf{b})|\Theta_\textbf{ab}\rangle\otimes |\Phi^{A_\text{rest}}\rangle |\Phi^{B_\text{rest}}\rangle\ ,
\eeq
where 
\beq
|\Theta_\textbf{ab}\rangle=\frac{|00\rangle_\textbf{ab}+|11\rangle_\textbf{ab}}{\sqrt{2}}
\eeq
is a maximally entangled state of a qubit $\textbf{a}$ in region $A$ and a qubit $\textbf{b}$ in region $B$,   $|\Phi^{A_\text{rest}}\rangle $ is the  state of the rest of qubits in region $A$, $|\Phi^{B_\text{rest}}\rangle $ is the state of the rest of qubits in region $B$, and $H_\textbf{b}$ is the Hadamard operator acting on qubit $\textbf{b}$.  

Therefore, we conclude that in this example state $\Omega^{(AB|C)}(\Psi_\text{clus})$ defined in Eq.(\ref{state}) is equal to
\begin{align}
&\frac{1}{4}\sum_{r_o,r_e=0}^1 |r_or_e\rangle\langle r_or_e|^{K_C}\otimes \text{tr}_{\overline{AB}}(\Pi^{(C)}_{r_0 r_1} |\Psi_\text{clus}\rangle\langle\Psi_\text{clus}|)\nonumber\\ &\cong
\frac{1}{4}\sum_{r_o,r_e=0}^1 |r_or_e\rangle\langle r_or_e|^{K_C}\otimes  |\Phi^{(AB)}_{r_or_e}\rangle\langle\Phi^{(AB)}_{r_or_e}| \ \ ,
\end{align}
where $\cong$ means the equality holds up to local unitaries on systems $A$ and $B$, which do not change the entanglement.   Next, note that because state $|\Phi^{A_\text{rest}}\rangle |\Phi^{B_\text{rest}}\rangle$ is uncorrelated across $A$ and $B$, the entanglement of  state $\Omega^{(AB|C)}(\Psi_\text{clus})$  relative to $A|K_C B$ partition, or  $AK_C| B$ partition, is equal to the entanglement of state 
\beq
\frac{1}{4}\sum_{r_o,r_e=0}^1 |r_or_e\rangle\langle r_or_e|^{K_C}\otimes  (I_\textbf{a}\otimes U_{r_or_e})|\Theta\rangle\langle\Theta|_\textbf{ab} (I_\textbf{a}\otimes U^\dag_{r_or_e})
\eeq   
where $U_{r_or_e}=X_\textbf{b}^{r_o}Z_\textbf{b}^{r_e}H_\textbf{b}$. As we have seen in Sec.(\ref{Sec:exa1}), this state can be reversibly transformed to state $|\Theta\rangle_\textbf{ab} $ via LOCC, and hence for any measure of entanglement $E$, its entanglement relative to $A|K_C B$ partition or  $AK_C| B$ partition is equal to the entanglement of state $|\Theta_\textbf{ab}\rangle$, i.e.
\beq
E(\Omega^{(AB|C)}(\Psi_\text{clus}))=E(|\Theta_\textbf{ab}\rangle) .
\eeq   
For instance, if we use the negativity $\mathcal{N}(\sigma^{AB})\equiv \frac{\|{\sigma^{AB}}^{\textbf{T}_{A}}\|_{1}-1}{2}$, as a measure of entanglement, then the above equation yields: 
\beq
\frac{1}{2}\sum_{r_0 r_1=0}^1\left\| \left[\text{tr}_{\overline{AB}}(\Pi^{(C)}_{r_0 r_1} |\Psi_\text{clus}\rangle\langle\Psi_\text{clus}|)\right]^{\textbf{T}_{A}}   \right\|_{1}-\frac{1}{2}=\frac{1}{2} \ .
\eeq
Note that this equality holds regardless of the size of regions $A$, $B$ and $C$ (as long as they contain at least two qubits).  

In conclusion, we see that this example is consistent with our general results: For the non-trivial phase of $\mathbb{Z}_2\times \mathbb{Z}_2$, the dimension of the irreducible projective representations is $d_{[\omega]}=2$, and therefore the SPT-Entanglement should be equal to the entanglement of a maximally entangled state of a pair of qubits, which is exactly the case for the cluster state.

In the following, we present a  general argument to show that why after measuring the charge in region $C$ on the cluster state, the regions $A$ and $B$ should become entangled. This argument is based on some symmetries of the cluster Hamiltonian, and hence applies to a more general class of perturbed cluster Hamiltonians.

\subsection{A general argument based on the symmetries of the Hamiltonian}
Let $C$ be a connected region on the ring with even number of qubits. Let qubits $\textbf{a}$ and  $\textbf{a}'$ be, respectively,  the first and the second qubits immediately outside $C$ in the left-hand side of this region, and qubits $\textbf{b}$ and $\textbf{b}'$  be, respectively,  the first and the second qubits immediately  outside $C$ in the right-hand side of this region. Define the unitaries
\bes\label{EQeven2}
\begin{align}
F^{(C)}_\text{odd}&=  (I_{\textbf{a}'}\otimes Z_{\textbf{a}}) \otimes X_\text{odd}^{(C)}\otimes (X_{\textbf{b}}\otimes Z_{\textbf{b}'}) \ ,\\  
F^{(C)}_\text{even}&= (Z_{\textbf{a}'}\otimes X_\textbf{a}) \otimes X_\text{even}^{(C)}\otimes (Z_{\textbf{b}}\otimes I_{\textbf{b}'}) \ ,
\end{align}
\ees
where $I$, $X$, $Y$, and $Z$ are, respectively, the qubit identity operator, and Pauli operators.  Note that 
\beq
[F^{(C)}_\text{even},F^{(C)}_\text{odd}]=[F^{(C)}_\text{odd}, H_\text{clus}]=[F^{(C)}_\text{even}, H_\text{clus}]=0 \ .
\eeq
These equations follow from the fact that  $F^{(C)}_\text{even}$ and $F^{(C)}_\text{odd}$ can be written as  products of $K_{i}$ operators, namely
\begin{align}
F^{(C)}_\text{even}&=K_\textbf{a}  
\Big[\prod_{\substack{i\in C\\
                  i: \text{even} }} K_i\Big]\  , \text{and} \ \ \
F^{(C)}_\text{odd}&=\Big[
\prod_{\substack{i\in C\\
                  i: \text{odd} }} K_i\Big] K_\textbf{b}\ .
\end{align}
Then, because all $K_i$ operators  commute with each other, it follows that $H_\text{clus}=- \sum_i K_i$ commutes with $F^{(C)}_\text{odd}$ and $F^{(C)}_\text{even}$, and  therefore these unitaries are symmetries of the cluster Hamiltonian.

As we show in Sec.(\ref{proofprop}), all Hamiltonians with such symmetries behave similar  to the cluster Hamiltonian in the following sense:
\begin{proposition}\label{prop49}
Suppose a Hamiltonian $H$ is invariant under unitaries $F^{(C)}_\text{even}$ and $F^{(C)}_\text{odd}$ defined in Eqs.(\ref{EQeven2}), i.e. $[H,F^{(C)}_\text{even}]=[H,F^{(C)}_\text{odd}]=0$. Then, given any non-degenerate eigenstate of this Hamiltonian, by measuring the observables $X_\text{even}^{(C)}$ and $X_\text{odd}^{(C)}$, and performing a unitary transformation on qubits $\textbf{aa}'$ and a unitary transformation on $\textbf{bb}'$ we can transform qubits $\textbf{a}$ and $\textbf{b}$ to a maximally entangled state. 
 \end{proposition}
More precisely, we show that given a non-degenerate eigenstate $|\Psi\rangle$ of Hamiltonian $H$,  if the outcome of $X_\text{even}^{(C)}$  measurement is $(-1)^{r_e}$ for $r_e \in\{0,1\}$ and the outcome of $X_\text{odd}^{(C)}$  measurement is $(-1)^{r_o}$ for $r_o \in\{0,1\}$, then by applying a \emph{Controlled-Z} unitary (defined below) on 
$\textbf{aa}'$ and a Controlled-Z unitary on $\textbf{bb}'$ we can transform the post-measurement state of system to 
\beq\label{rty}
|\Phi_{r_or_e}\rangle= (I_\textbf{a}\otimes X_\textbf{b}^{r_o+f_o}Z_\textbf{b}^{r_e+f_e}H_\textbf{b})|\Theta_\textbf{ab}\rangle\otimes |\Phi_{\overline{\textbf{ab}}}\rangle\ ,
\eeq
where  $|\Phi_{\overline{\textbf{ab}}}\rangle$ is the state of all qubits except $\textbf{a}$ and $\textbf{b}$, and $H_\textbf{b}$ is the Hadamard operator acting on qubit $\textbf{b}$. Here, the two bits $f_o,f_e\in\{0,1\}$ are defined by $(-1)^{f_e}=\langle\Psi|F^{(C)}_\text{even}|\Psi\rangle$ and $(-1)^{f_o}=\langle\Psi|F^{(C)}_\text{odd}|\Psi\rangle$. Finally, the controlled-Z unitary is the two-qubit unitary which maps $|11\rangle$ to $-|11\rangle$ and leaves states $|00\rangle$, $|01\rangle$, and $|10\rangle$  unchanged.

Note that  state in Eq.(\ref{cluster-ent}), which is obtained from the cluster state, is a special case of state in Eq.(\ref{rty})(For cluster state $f_e=f_o=0$). 

Proposition \ref{prop49} has the following corollary: Suppose we add a perturbation to the cluster Hamiltonian $H_\text{clus}$, which (i) acts trivially on qubits $\textbf{a}$, $\textbf{a}'$,  $\textbf{b}$, and $\textbf{b}'$, and (ii) commutes with $X_\text{even}^{(C)}$ and $X_\text{odd}^{(C)}$. Then, if an eigenstate of the perturbed Hamiltonian is  non-degenerate,  by measuring $X_\text{even}^{(C)}$ and $X_\text{odd}^{(C)}$, and applying local unitaries on qubits $\textbf{aa}'$ and $\textbf{bb}'$, we can transform qubits $\textbf{a}\textbf{b}$ to a maximally entangled state.

This corollary implies that the SPT-Entanglement for non-degenerate eigenstates of such Hamiltonians is non-zero regardless of the size of region $C$, and therefore, it follows from our results on SPT-Entanglement that the state should be in the non-trivial SPT  phase. That is it cannot be transformed to a product state by a low-depth circuit which respects $\mathbb{Z}_2\times \mathbb{Z}_2$ symmetry represented by $\bigotimes_{i\in \text{even}} X_i$ and $\bigotimes_{i\in \text{odd}} X_i$.

For instance, suppose we perturb the cluster Hamiltonian by adding a (possibly non-uniform) magnetic field in $x$ direction everywhere throughout the system,  except on the four qubits   $\textbf{a}\textbf{a}'$,  and $\textbf{b}\textbf{b}'$, where the distance between $\textbf{a}\textbf{a}'$ and $\textbf{b}\textbf{b}'$ is arbitrary large, i.e. of the order of the system size. Then, our result implies that if  the ground state of such Hamiltonian is non-degenerate, then it should be in the non-trivial SPT phase. More specifically, we know that  by measuring operators $X_\text{even}^{(C)}$ and $X_\text{odd}^{(C)}$ and applying controlled-Z  unitaries on  $\textbf{aa}'$ and $\textbf{bb}'$, we can transform qubits $\textbf{a}\textbf{b}$ to a maximally entangled state, similar to the case of cluster state.

Note that the maximally entangled state we obtain in this way might be different from the one we obtain from the cluster state. This is because the final state of $\textbf{a}$ and  $\textbf{b}$, not only depends on the outcomes of $X_\text{even}^{(C)}$ and $X_\text{odd}^{(C)}$  measurements (encoded in $r_o$ and $r_e$), but also depends on the value of the bits $f_e$ and $f_o$.   This fact has important consequences in the context of MBQC. We discuss more about this in Sec.(\ref{Sec:iden}) 


In the above argument  we assumed the perturbation acts trivially on the four qubits  $\textbf{a}\textbf{a}'$, and $\textbf{b}\textbf{b}'$. Clearly this assumption does not hold if, for example, we apply a uniform magnetic field in $x$ direction to all qubits in the system, that is if we add the term $B\sum_i X_i$ to $H_\text{clus}$.  But, in this situation we can decompose the perturbation to two parts: a part which acts trivially on  these four qubits, and a small part which acts non-trivially on them. In the example of uniform magnetic field in $x$ direction, for instance, we can write
\bes
\begin{align}
H_\text{clus}(B)&=H_\text{clus}+B \sum_i X_i\\ &=\tilde{H}_\text{clus}+B(X_\textbf{a}+X_{\textbf{a}'}+X_{\textbf{b}}+X_{\textbf{b}'} )\ ,
\end{align}
\ees
where $\tilde{H}_\text{clus}=H_\text{clus}+B \sum_{i\neq \textbf{a},\textbf{a}',  \textbf{b}, \textbf{b}'} X_i$. Clearly, $\tilde{H}_\text{clus}$ satisfies the symmetry conditions required by  the above proposition. Then, it follows form the proposition that if the ground state of  $\tilde{H}_\text{clus}$ is unique then it should have a nonzero SPT-Entanglement, regardless of the size of  region $C$, and  therefore, using our general results on SPT-Entanglement, we conclude that it should be  in the non-trivial SPT phase.

Now we can think of  the term $B(X_\textbf{a}+X_{\textbf{a}'}+X_{\textbf{b}}+X_{\textbf{b}'} )$ as a small (norm-bounded) symmetric perturbation on the Hamiltonian $\tilde{H}_\text{clus}$.  Assume we know the ground state of $H_\text{clus}(B)$ is non-degenerate, and its energy gap with the first excited state is $\Delta E(B)$. Then, if $ \Delta E(B)>8|B|$, we can conclude  that the ground state of Hamiltonian $\tilde{H}_\text{clus}$ is also non-degenerate, and has an energy gap larger than $ \Delta E(B)-8|B|$ (This follows from the fact that adding a perturbation $V$ to a Hamiltonian can change the energy of each eigenstate by, at most, $\|V\|$, where $\|\cdot\|$ is the largest singular value of the operator). Indeed, considering the family of Hamiltonians $\tilde{H}_\text{clus}+s B(X_\textbf{a}+X_{\textbf{a}'}+X_{\textbf{b}}+X_{\textbf{b}'} )$ for $s\in[0,1]$,  we can easily see that if $8|B|< \Delta E(B)$, then there exists a smooth path in the space of gapped symmetric local Hamiltonians which connects $\tilde{H}_\text{clus}$ to $H_\text{clus}(B)$. This implies that the ground state of $\tilde{H}_\text{clus}$ and $H_\text{clus}(B)$ should be in the same SPT phase. More precisely,  the ground state of the Hamiltonian $\tilde{H}_\text{clus}+B(X_\textbf{a}+X_{\textbf{a}'}+X_{\textbf{b}}+X_{\textbf{b}'} )$ can be obtained from the ground state of $\tilde{H}_\text{clus}$ via  a low-depth symmetric circuit acting in a neighborhood around $\textbf{aa}'$  and $\textbf{bb}'$. 

To summarize, we conclude that if ${H}_\text{clus}(B)$ has a unique ground state with gap $\Delta E(B)>8|B|$ then it should be in the non-trivial SPT phase protected by $\mathbb{Z}_2\times \mathbb{Z}_2$. This is consistent with the previously known result in \cite{Doherty2009}.

 
We can easily generalize this argument for a general perturbation  $\sum_j V_j$, which is sum of symmetric local terms $V_j$. The idea is that we choose a pair of neighbor qubits $\textbf{a}\textbf{a}'$ and another pair of neighbor qubits $\textbf{b}\textbf{b}'$ which are arbitrary far from each other. 
Then, we adiabatically turn off all the terms $V_j$  which act non-trivially on each of the qubits $\textbf{a}$, $\textbf{a}'$, $\textbf{b}$ and $\textbf{b}'$. We also turn off all the terms which act non-trivially on both sides of $\textbf{aa}'$, i.e. those which connect the qubits on the left side of $\textbf{aa}'$ to the ones on the right side of $\textbf{aa}'$. Similarly, we turn of all the interactions which act on both sides of $\textbf{bb}'$.  Then, if the perturbation $\sum_j V_j$ is local  and the gap of Hamiltonian $H=H_\text{clus}+\sum_j V_j$ is large enough compared to the norm of $V_j$ operators, as we turn off these interactions the gap does not vanish. This means that the modified Hamiltonian $\tilde{H}$ is connected to the original Hamiltonian $H=H_\text{clus}+\sum_j V_j$ via a smooth path of gapped symmetric Hamiltonians, and therefore they are in the same SPT phase. Furthermore, if the ground state of $H$ is  unique then the ground state of $\tilde{H}$ should also be unique. 

Then, using the facts that (i) all $V_j$ operators commute with both $\bigotimes_{i\in \text{even}} X_i$ and $\bigotimes_{i\in \text{odd}} X_i$, and (ii) all the remaining $V_j$ operators  in $\tilde{H}$ acts trivially on $\textbf{aa}'$ and  $\textbf{b}'\textbf{b}$, and they do not connect the two sides of these qubits, we can easily see that $[\tilde{H},F^{(C)}_\text{even}]=[\tilde{H},F^{(C)}_\text{odd}]=0$, where $C$ is the region between $\textbf{aa}'$ and  $\textbf{b}'\textbf{b}$, and for simplicity we assume it has even number of qubits. 

 
 Therefore, because $[\tilde{H},F^{(C)}_\text{even}]=[\tilde{H},F^{(C)}_\text{odd}]=0$, and $\tilde{H}$ has a unique ground state,   by proposition \ref{prop49} we know that by measuring charge in region $C$, one can create a maximally entangled state between $a$ and $b$, which implies the state should be in the non-trivial SPT phase. Since $H$ and $\tilde{H}$ are in the same SPT phase, this implies the ground state of the original Hamiltonian $H$ should also be in the non-trivial SPT phase.
 
Using this argument we can easily prove the following result:
\begin{theorem}
Consider the perturbed cluster Hamiltonian $H=H_\text{clus}+\sum_j V_j$, where each term $V_j$  has $\mathbb{Z}_2\times \mathbb{Z}_2$ symmetry (i.e. commutes with both $\bigotimes_{i\in \text{even}} X_i$ and $\bigotimes_{i\in \text{odd}} X_i$), and is $k$-local (i.e. it acts non-trivially, on at most $k$ neighbor qubits). Also, assume there are, at most, $t$ different terms $V_j$ which act non-trivially on any given qubit in the system. Then, if the ground state of $H$ is non-degenerate and has the spectral gap $\Delta E > 4 t(k+1) \times \max_j \|V_j\|$, then it is in the  non-trivial SPT phase protected by $\mathbb{Z}_2\times \mathbb{Z}_2$ symmetry.
\end{theorem}
We can also directly argue that for the ground state of such Hamiltonians, the SPT-Entanglement  between two sufficiently large regions $A$ and $B$ should be nonzero, regardless of the distance between $A$ and $B$: Suppose we choose the connected region $A$  such that  qubits $ \textbf{a}\textbf{a}'$  are somewhere in the middle of region $A$,  far from its boundaries. Similarly, we choose the connected region  $B$ such that  $ \textbf{b}\textbf{b}'$ are in this region, and far from its boundaries. Then, from  \ref{prop49} we know that if in the ground state of $\tilde{H}$, we measure the charge in the region between qubits  $ \textbf{a}\textbf{a}'$ and $ \textbf{b}\textbf{b}'$ then the entanglement between  $ \textbf{a}\textbf{a}'$ and $ \textbf{b}\textbf{b}'$  is nonzero. 

Now as we add the extra terms  to $\tilde{H}$, which act non-trivially on  $ \textbf{a}\textbf{a}'$ and $ \textbf{b}\textbf{b}'$, because the system is gapped,  the effect of  these local perturbations on the ground state is equivalent to two local (symmetric) unitaries  which act in a neighborhood around  $ \textbf{a}\textbf{a}'$ and $ \textbf{b}\textbf{b}'$. In other words, there are local unitaries $V_{A}$ and $V_{B}$ which act respectively, in a neighborhood around  $ \textbf{a}\textbf{a}'$ and  $ \textbf{b}\textbf{b}'$, such that $V_{A} V_{B}$ transforms the ground state of  $\tilde{H}$ to the ground state of $H$. But, if regions $A$ and $B$ are large enough then the support of these unitaries are contained in these regions, and therefore applying these local unitaries cannot change  the SPT-Entanglement of state between regions $A$ and $B$. It follows that in the limit where regions $A$ and $B$ are large then the SPT-Entanglement should be nonzero, regardless of the distance between $A$ and $B$.

\subsection{Identity gate in MBQC is the charge measurement}\label{Sec:iden}

As we mentioned before, it has been shown that the computational power of the 1-dimensional cluster state can be  understood as a consequence of the fact that this state lies in the non-trivial SPT phase protected by the  $\mathbb{Z}_2\times \mathbb{Z}_2$ symmetry. More precisely, it has been argued by Else et. al \cite{Else2012,ElseBartlettDoherty} that for any other state of a 1-dimensional chain that lies in the same SPT phase, the operation which corresponds to the \emph{identity gate} in MBQC can be implemented over arbitrary long distances (See also \cite{Jac,Henry, Robert, Stephen2009}). Their  argument is based on the classification of the SPT phases in the MPS framework, and holds for a more general class of SPT phases, which correspond to the  \emph{maximally non-commutative} cohomology classes \cite{Else2012,ElseBartlettDoherty}.



It is interesting to note that the measurements required to implement the identity gate on the 1-dimensional cluster state are basically equivalent to a charge measurement:  to implement the identity gate one measures  $X$ operators on all qubits between the two endpoints, and then finds  the parity of the outcomes of measurements on the even and odd qubits. As we have seen before, this process basically measures the total charge corresponding to  $\mathbb{Z}_2\times\mathbb{Z}_2$ symmetry in the region between the two qubits. This simple observation may, to some extent, demystify the robustness of the computational power of 1-dimensional cluster state under symmetric perturbations, and provide a new insight into this important result.  For instance, consider the cluster state on an open chain, and suppose we apply an arbitrary symmetric unitary which acts trivially on the two qubits at the two endpoints of the chain.  Under any such unitary the total charge in the region between the two endpoints remains conserved. This immediately implies that the transformed  state can also be used to implement the identity gate between the two endpoints, exactly in the same way that the cluster state itself can be used. 


Finally, it is worth noting  that  the fact that the state is in the non-trivial SPT phase alone, does not mean that we can use it as a resource for quantum computation or teleportation.  For example, consider the family of Hamiltonians 
\beq
H_\text{clus}(\theta)=S(\theta)H_\text{clus} S^\dag(\theta)\ ,
\eeq
which are obtained from the cluster Hamiltonian by applying the local unitary
\beq
S(\theta)=\cos \theta I+i \sin \theta (Z_{j-1}\otimes Z_{j+1})\ ,
\eeq
where $j$ is an arbitrary qubit in the system. Observe that the unitary $S(\theta)$ commutes with both   $\bigotimes_{i\in \text{even}} X_i$ and $\bigotimes_{i\in \text{odd}} X_i$, and therefore all the Hamiltonians in this family have  $\mathbb{Z}_2\times\mathbb{Z}_2$ symmetry. Furthermore, all these Hamiltonians have the same spectrum as the cluster Hamiltonian $H_\text{clus}$, and hence they are gapped. It follows that for any value of $\theta$ there is a smooth path of symmetric gapped Hamiltonians which connects $H_\text{clus}(\theta)$ to $H_\text{clus}$, and therefore their ground states are in the same SPT phase (Equivalently, their ground states can be transformed to each other via a low-depth symmetric circuit, namely $S(\theta)$ itself).


Suppose we want to use the ground state of $H_\text{clus}(\theta)$ to perform the identity gate between qubit $j$ and another qubit  $k\neq j\pm 1$ in the system.  It can be easily shown that if we follow the same steps that we do in the case of cluster state, i.e. we measure the charge in the region between the two qubits and then decouple them from the rest of qubits by applying controlled-Z unitaries,  then we can still create a maximally entangled state between qubits $j$ and any arbitrary qubit  $k\neq j\pm 1$, regardless of their distance. However, the maximally entangled state  we create in this case is different from the one we obtain in the case of cluster state, by a local unitary which depends on $\theta$. Therefore, if we do not have any information about $\theta$, then the average state of the two qubits will be the totally mixed state, and hence it cannot be used for teleportation or quantum computation.

We conclude that to be able to exploit the computational power of the perturbed cluster Hamiltonian just knowing that the ground state is in the non-trivial SPT phase is not enough, and one needs to have further information about the perturbation, or equivalently about the ground state of the perturbed Hamiltonian. In particular, if one assumes the  perturbation is sufficiently weaker than the  gap of the cluster Hamiltonian, then the  maximally entangled state obtained from the ground state of the perturbed Hamiltonian is close to the one obtained from the cluster state, even if the two qubits $j$ and $k$ are arbitrary far from each other  (In the above example, this means that we need to assume $|\theta|\ll 1$).

\subsection{Proof of proposition \ref{prop49}}\label{proofprop}
In the rest of this section we present the proof of proposition \ref{prop49}.

Let $|\Psi\rangle$  be a non-degenerate eigenstate of $H$. The fact that  $F^{(C)}_\text{odd}$ and $F^{(C)}_\text{even}$ commute with $H$, together with the fact that this eigenstate is non-degenerate implies that $|\Psi\rangle$ should also be an eigenstate of $F^{(C)}_\text{odd}$ and $F^{(C)}_\text{even}$. Using the facts that $F^{(C)}_\text{odd}$ and $F^{(C)}_\text{even}$ are both unitary operators, and the square of both is the identity, we find that 
\bes
\begin{align}
F^{(C)}_\text{odd}|\Psi\rangle=(-1)^{f_\text{o}} |\Psi\rangle\ ,\\ 
F^{(C)}_\text{even}|\Psi\rangle=(-1)^{f_\text{e}} |\Psi\rangle\ ,
\end{align}
\ees
where $f_\text{o},f_\text{e}\in\{0,1\}$.

Next, note that unitaries $F^{(C)}_\text{odd}$ and $F^{(C)}_\text{even}$ commute with  $X_\text{even}^{(C)}$ and $X_\text{odd}^{(C)}$, and hence they commute with 
\beq\label{eq:copy}
\Pi^{(C)}_{r_or_e}=\big(\frac{I+(-1)^{r_o}X_\text{odd}^{(C)}}{2}\big) \big(\frac{I+(-1)^{r_e}X_\text{even}^{(C)}}{2}\big)\ .
\eeq
This immediately implies that the vectors $\Pi_{r_or_e}^{(C)}  |\Psi\rangle$ are also eigenstates of $F^{(C)}_\text{odd}$ and $F^{(C)}_\text{even}$ with eigenvalues $\pm 1$, 
\bes\label{wed}
\begin{align}
F^{(C)}_\text{odd} \Pi_{r_or_e}^{(C)} |\Psi\rangle&=\Pi_{r_or_e}^{(C)} F^{(C)}_\text{odd} |\Psi\rangle=(-1)^{f_\text{o}}  \Pi_{r_or_e}^{(C)}  |\Psi\rangle\\  F^{(C)}_\text{even} \Pi_{r_or_e}^{(C)} |\Psi\rangle&=\Pi_{r_or_e}^{(C)} F^{(C)}_\text{even} |\Psi\rangle=(-1)^{f_\text{e}}  \Pi_{r_or_e}^{(C)}  |\Psi\rangle\  .
\end{align}
\ees
Next, observe that  $\Pi_{r_or_e}^{(C)}X_\text{odd}^{(C)}=(-1)^{r_o} \Pi_{r_or_e}^{(C)}$, and $\Pi_{r_or_e}^{(C)}X_\text{even}^{(C)}=(-1)^{r_e} \Pi_{r_or_e}^{(C)}$, which can be seen using Eq.(\ref{eq:copy}). Therefore, multiplying both sides of 
\begin{align*}
F^{(C)}_\text{odd}&=  (I_{\textbf{a}'}\otimes Z_{\textbf{a}}) \otimes X_\text{odd}^{(C)}\otimes (X_{\textbf{b}}\otimes Z_{\textbf{b}'})\\ F^{(C)}_\text{even}&= (Z_{\textbf{a}'}\otimes X_\textbf{a}) \otimes X_\text{even}^{(C)}\otimes (Z_{\textbf{b}}\otimes I_{\textbf{b}'})\ ,
\end{align*}
in $\Pi_{r_or_e}^{(C)}$ we find
\bes
\begin{align}
F^{(C)}_\text{odd}\Pi_{r_or_e}^{(C)}&=(-1)^{r_o}  (Z_{\textbf{a}}\otimes X_{\textbf{b}}\otimes Z_{\textbf{b}'}) \Pi_{r_or_e}^{(C)}\ , \\
F^{(C)}_\text{even}\Pi_{r_or_e}^{(C)}&=(-1)^{r_e}(Z_{\textbf{a}'}\otimes X_\textbf{a}\otimes Z_{\textbf{b}}) \Pi_{r_or_e}^{(C)}    \ .
\end{align}
\ees
This together with Eqs.(\ref{wed}) imply that
\bes\label{qdf}
\begin{align}
\left[Z_{\textbf{a}}\otimes (X_{\textbf{b}}\otimes Z_{\textbf{b}'})\right] \Pi_{r_or_e}^{(C)}|\Psi\rangle &=(-1)^{r_o+f_o}  \Pi_{r_or_e}^{(C)}|\Psi\rangle\ , \\
\left[(Z_{\textbf{a}'}\otimes X_\textbf{a})\otimes Z_{\textbf{b}}\right]\Pi_{r_or_e}^{(C)}|\Psi\rangle &=(-1)^{r_e+f_e}\Pi_{r_or_e}^{(C)}|\Psi\rangle   \ .
\end{align}
\ees

Let CZ be the controlled-Z unitary, i.e. the two-qubit unitary which maps $|11\rangle$ to $-|11\rangle$ and leaves states $|00\rangle$, $|01\rangle$, and $|10\rangle$  unchanged. Define 
\beq
|\Phi_{r_or_e}\rangle= \frac{[CZ_{{\textbf{a}}{\textbf{a}'}}\otimes CZ_{{\textbf{b}}{\textbf{b}'}}]\Pi_{r_or_e}^{(C)}|\Psi\rangle}{\sqrt{\langle\Psi|\Pi_{r_or_e}^{(C)}|\Psi\rangle}}\ ,
\eeq
to be the  state of system after projecting region $C$ to the sector with charge $r_or_e$, and applying a controlled-Z unitary on qubits ${\textbf{a}}$ and ${\textbf{a}'}$ and a controlled-Z unitary to qubits ${\textbf{b}}$ and ${\textbf{b}'}$. Then, by applying the operator $CZ_{{\textbf{a}}{\textbf{a}'}}\otimes CZ_{{\textbf{b}}{\textbf{b}'}}$ on both sides of Eqs.(\ref{qdf}), and using the fact that $CZ(X\otimes Z)CZ=(X\otimes I)$, we find
\bes
\begin{align}
\left(Z_{\textbf{a}}\otimes X_{\textbf{b}}\right) |\Phi_{r_or_e}\rangle &=(-1)^{r_o+f_o}  |\Phi_{r_or_e}\rangle\ , \\
\left(X_\textbf{a}\otimes Z_{\textbf{b}}\right)|\Phi_{r_or_e}\rangle &=(-1)^{r_e+f_e}|\Phi_{r_or_e}\rangle \ .
\end{align}
\ees
It can be easily shown that these two equations together uniquely determine the state of qubits  ${\textbf{a}}$ and ${\textbf{b}}$.  More precisely, up to a global phase, the only state of qubits ${\textbf{a}}$ and ${\textbf{b}}$ which satisfies these  equations is  the state
\beq
(I_\textbf{a}\otimes X_\textbf{b}^{r_o+f_o}Z_\textbf{b}^{r_e+f_e}H_\textbf{b})|\Theta_\textbf{ab}\rangle\ ,
\eeq
where $|\Theta_\textbf{ab}\rangle=\frac{|00\rangle_\textbf{ab}+|11\rangle_\textbf{ab}}{\sqrt{2}}$,  and $H_\textbf{b}$ is the Hadamard gate. It follows that any state $|\Phi_{r_or_e}\rangle$ which satisfies these equations should be in the form of
\beq
|\Phi_{r_or_e}\rangle= (I_\textbf{a}\otimes X_\textbf{b}^{r_o+f_o}Z_\textbf{b}^{r_e+f_e}H_\textbf{b})|\Theta_\textbf{ab}\rangle\otimes |\Phi_{\overline{\textbf{ab}}}\rangle\ ,
\eeq
where  $ |\Phi_{\overline{\textbf{ab}}}\rangle$ is  the state of the rest of qubits in the system, which is arbitrary (i.e. it is not constrained by these equations). This completes the proof of the proposition.




\section{SPT-Entanglement in 1-D systems}\label{Sec:1-d}
 In this section, we use the classification of the SPT phases in the Matrix Product State (MPS)  framework to calculate SPT-Entanglement in the case of 1-dimensional systems. We start with a short review of this classification (we follow the presentation of \cite{Schuch-Cirac2} and \cite{Else2013}).   

\subsection{MPS representation of SPT phases}

\begin{figure} [h]
\begin{center}
\includegraphics[scale=.39]{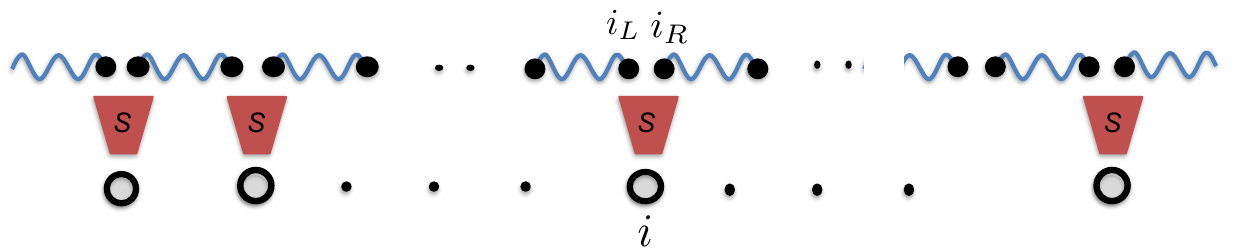}
\caption{Isometry $S$ maps virtual subsystems $i_L$ and $i_R$ to the physical site $i$. 
}\label{MPS}
\end{center}
\end{figure}

For a 1-dimensional system with short range correlations by blocking the sites in the large blocks a translationally invariant ground state will converge to a fixed point in the form of
\begin{equation}\label{def-psi}
|\Psi\rangle=S^{\otimes N} |\lambda\rangle^{\otimes N}\ ,
\end{equation}
where  $ |\lambda\rangle= \sum_{k} \lambda_k |k\rangle|k\rangle$ is a \emph{virtual} entangled state between adjacent virtual sites   with Schmidt coefficients $\{\lambda_{k}\}$, and $S$ is an isometry which  maps two virtual sites $i_{L}$ and $i_{R}$ to a physical site $i$ (See Fig.\ref{MPS}) \cite{Cirac-RG, Cirac-MPS}. This isometry can indeed be thought as a Renormalization Group (RG) transformation  on a block \cite{Cirac-RG, Cirac-MPS}. It is shown that for any gapped 1-dimensional system with a unique ground state, by blocking the physical sites, the ground state can be approximated by a state in the form of $|\Psi\rangle$ with an accuracy which is exponential in the block sizes \cite{Cirac-RG, Cirac-MPS}. 

Consider a symmetry group $G$ with on-site linear unitary representation $g\rightarrow u(g)$.  After blocking $L$ sites together  this symmetry will be represented by  $g\rightarrow U(g)$ on each block, where $U(g)=u(g)^{\otimes L}$. Assume $|\Psi\rangle$ is invariant under this symmetry, i.e. $U(g)^{\otimes N}|\Psi\rangle=|\Psi\rangle$.  Then, it turns out that there always exists a projective representation $g\rightarrow V(g)$ of group $G$ such that 
\beq
U_{i}(g)S=S \left[V_{i_L}(g)\otimes {V}^{\ast}_{i_R}(g) \right]
\eeq
and
\beq
 \left[{V}_{i_R}^{\ast}(g)\otimes V_{(i+1)_{L}}(g)\right] |\lambda\rangle=|\lambda\rangle\ ,
\eeq 
where $V^{\ast}(g)$ is the complex conjugate of $V(g)$ \cite{Cirac-String}. Note that for any phase $e^{i\theta(g)}$ the representation $g\rightarrow e^{i\theta(g)}V(g)$ will also satisfy the above equations, and hence $V(g)$ is defined only up to a phase. Let $\omega$ be the 2-cocycle of the representation $g\rightarrow V(g)$, i.e. 
\beq
\forall g,h\in G:\ \ \ V(g)V(h)=\omega(g,h) V(gh)\ .
\eeq
Then, the gauge transformation $V(g)\rightarrow e^{i\theta(g)}V(g)$ induces an equivalence relation on the space of the 2-cocycles: $\omega$ and $\omega'$ are equivalent, or belong to the same cohomology class, if there exists a phase $e^{i\theta(g)}$ for which
\beq
\forall g,h\in G:\ \ \  \omega(g,h)=\omega'(g,h)e^{i[\theta(gh)-\theta(g)-\theta(h)]}\ .
\eeq
For any 2-cocycle $\omega$, its corresponding equivalence class is denoted by $[\omega]$.  The set of equivalence classes forms a group, which is called the second cohomology class of $G$, and is denoted by $H^{2}(G,U(1))$. 

 According to the classifications of the SPT phases of 1-dimensional systems,  the equivalence class $[\omega]$ uniquely determines the SPT phase  \cite{Clas-Wen,Clas-Wen2,Pollman-Turner2, Schuch-Cirac}. For example, the 2-cocycles corresponding  to the Abelian group $\mathbb{Z}_2\times \mathbb{Z}_2$  have two different equivalence classes, which correspond to two different phases protected by this symmetry.  See Sec. \ref{Sec:Example} for further discussion about the example of $\mathbb{Z}_2\times \mathbb{Z}_2$.
  

\subsection{Cohomology of Abelian groups}

Interestingly, it turns out that for \emph{Abelian} groups the irreducible projective representations  in the same cohomology class all have the same dimension. Indeed, in the  Appendix we prove that
\begin{lemma}\label{lem-new-dim}
Let $g\rightarrow u_\alpha(g)$ and $g\rightarrow u_\beta(g)$  be two finite-dimensional projective irreducible representations of an Abelian group $G$, whose 2-cocycles  belong to the same cohomology class. Then, these representations have the same dimensions. Furthermore,  there exists a unitary $W$  and a phase $e^{i r(g)}$ such that
\beq
\forall g\in G:\ \ u_\beta(g)=e^{i r(g)} Wu_\alpha(g) W^{\dag}\ .
\eeq
\end{lemma}
Basically, this lemma means that in the case of Abelian groups,  up to a unitary and a phase freedom, there is only a unique irreducible projective representation in each cohomology class. Note that this is not true for a general non-Abelian group.

In the following, the dimension of the projective irreducible representations in the cohomology $[\omega]$ is denoted by $d_{[\omega]}$.

\subsection{Main result in 1-dimensional systems}
We calculate the SPT-Entanglement for states in the form of state in Eq.(\ref{def-psi}), that is for the fixed points of the RG.  For a general MPS, which is not in this form, we know that by blocking the sites, the state converges to a  fixed point in the form of  Eq.(\ref{def-psi}).  Therefore, for a general MPS if we calculate SPT-Entanglement for large regions  $A$, $B$ and $C$ the result should converge to what we obtain for these fixed points.  Note that  other approaches have been recently proposed  for detecting SPT order based on the properties of the fixed point of the RG \cite{Chen2,Singh}.

Let $A$ and $B$ be any two non-neighbor blocks of a 1-dimensional system, and $C$ be a (connected) region between $A$ and $B$.
\begin{theorem}\label{Thm}
Let  $|\Psi\rangle$ be  a state in the form of Eq.(\ref{def-psi}) in the SPT phase corresponding to the cohomology class $[\omega]$. Then, with respect to any measure of entanglement $E$, the SPT-Entanglement between  $A$ and $B$ is equal to the entanglement of a maximally entangled state of a pair of $d_{[\omega]}$-dimensional systems. 
\end{theorem}
In particular, if we use the negativity to quantify SPT-Entanglement, then this theorem implies
\beq
\sum_{\kappa}\left\| \left[\text{tr}_{\overline{AB}}(\Pi^{(C)}_{\kappa} |\Psi\rangle\langle\Psi|)\right]^{\textbf{T}_{A}}   \right\|_{1}= d_{[\omega]}\ ,
\eeq
which is equal to 1 in the trivial phase, and larger than or equal to 2 in the non-trivial SPT phases. Note that since the SPT phase of state $|\Psi\rangle$ is uniquely determined by the equivalence class $[\omega]$, this theorem implies that for states in the form of  Eq.(\ref{def-psi}),  the SPT-Entanglement depends only on the SPT phase of the state.

Consider state
\beq
\Omega^{(AB|C)}(\Psi)=\sum_{\kappa} |\kappa\rangle\langle\kappa|^{(K_{C})}\otimes \text{tr}_{\overline{AB}}(\Pi^{(C)}_{\kappa} |\Psi\rangle\langle\Psi|) \ ,
\eeq
and let
\beq
|\Theta\rangle \equiv \frac{1}{\sqrt{d_{[\omega]}}}\sum_{k=1}^{d_{[\omega]}} |kk\rangle\ 
\eeq
be a maximally entangled state of a pair of $d_{[\omega]}$-dimensional systems.
  Then, another way to phrase theorem \ref{Thm}  is that both  transformations $|\Theta\rangle \xrightarrow{\text{LOCC}} \Omega^{(AB|C)}({\Psi})$ and $\Omega^{(AB|C)}({\Psi})\xrightarrow{\text{LOCC}}|\Theta\rangle$ can be implemented via LOCC. In the following we prove this result.

\subsection{Finding SPT-Entanglement in MPS framework}
\begin{figure} [h]
\begin{center}
\includegraphics[scale=.33]{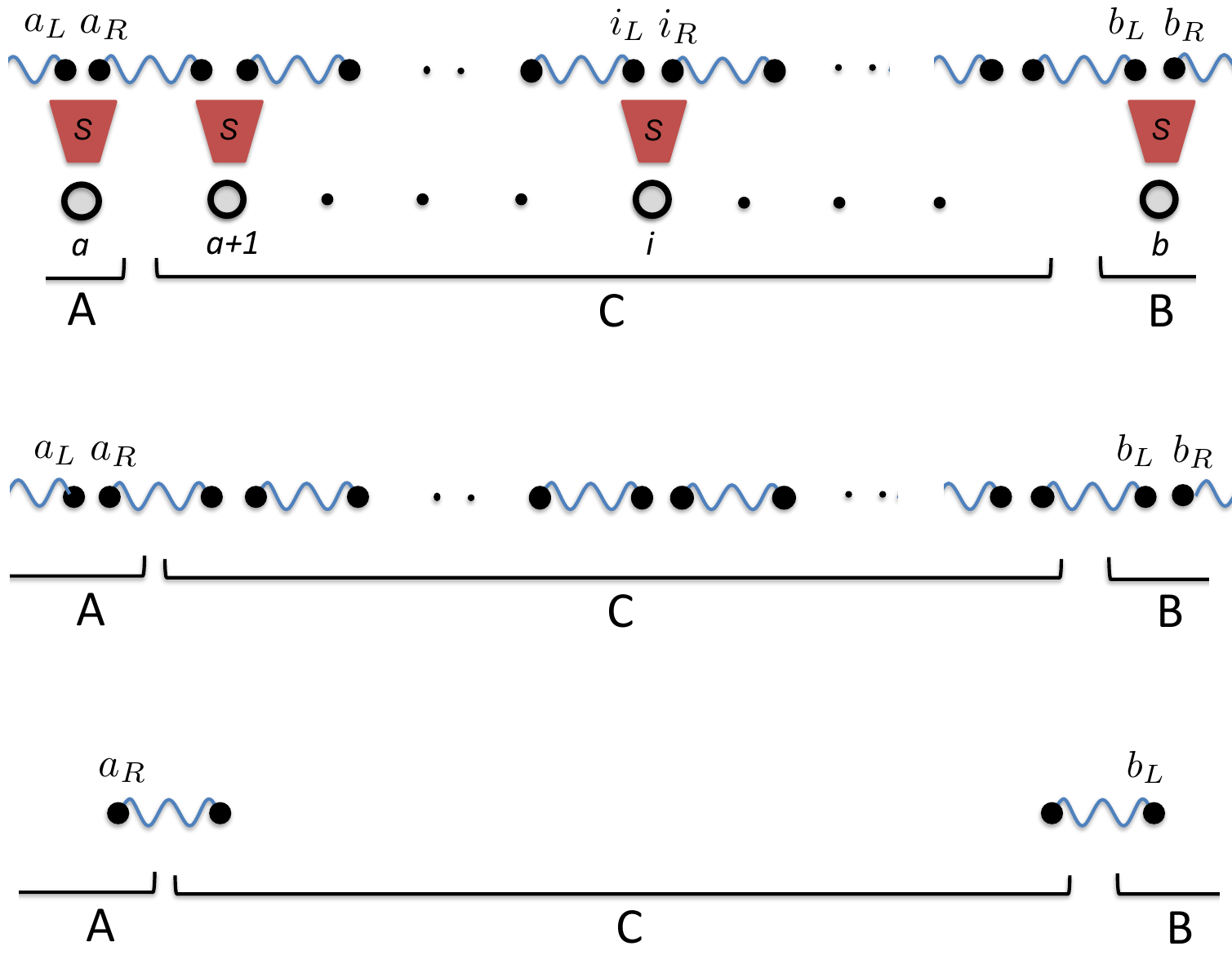}
\caption{Top: We  look at the entanglement of regions $A$ and $B$, given the value of charge in region $C$ is known. Here, each isometry $S$ maps two virtual systems to a physical system. Middle: Applying local isometry $S$ does not change entanglement, or charge. Therefore, we can ignore the isometries and look at the charge and entanglement at the level of the virtual systems. Bottom: For each pair of virtual systems in state $|\lambda\rangle$ the total charge is zero. Thus, if both pairs are inside region $C$ then they do not contribute  in the total charge in this region, and therefore we can ignore them. Furthermore,  all the virtual subsystems in $A$ or $B$, except  the  ones at the boundary, are uncorrelated with the virtual systems in the other regions. Therefore, since we are only interested in the entanglement between the two regions, we can ignore them. The resulting state is state $\sigma^{(ab)}$ defined in proposition \ref{prop23}, where $a$ and $b$ correspond,  respectively, to the virtual systems $a_R$ and $b_L$ in this figure. 
}\label{fig:MPS}
\end{center}
\end{figure}

The first step in the proof of theorem \ref{prop23} is the following simple observation, which is depicted in Fig.(\ref{fig:MPS}).
\begin{proposition}\label{prop23}
Applying local operations on Alice's and Bob's systems, we can reversibly transform state  $\Omega^{(AB|C)}(\Psi)$  to state 
\begin{align}\label{sigmadef}
\sigma^{(ab)}=\sum_\kappa |\kappa\rangle\langle\kappa|^{(K_C)}\otimes  \Tr_{{a'}{b'}}(\tilde{\Pi}^{({a'b'})}_{\kappa} \large[|\lambda\rangle\langle\lambda|_{aa'}\otimes|\lambda\rangle\langle\lambda|_{b'b} \large]) \nonumber\ ,
\end{align}
where  
\beq
\tilde{\Pi}^{({a'b'})}_{\kappa}=\frac{1}{|G|}\sum_{g\in G} e^{-i\kappa(g)}\  V_{a'}(g) \otimes V^\ast_{b'}(g)\ .
\eeq
\end{proposition}
This implies that state $\Omega^{(AB|C)}(\Psi)$ and state $\sigma^{(ab)}$ have exactly the same entanglement properties.  

Note that the set of projectors $\{\tilde{\Pi}^{({a'b'})}_{\kappa}: \kappa\in Q\}$ describes the charge measurement on the virtual systems  ${a}'$ and ${b}'$.  

This proposition follows from the following simple observations which are depicted in Fig.(\ref{fig:MPS}): (i) Applying local isometries do not change the entanglement of state. Therefore, instead of entanglement in the physical space, we can look at the entanglement in the virtual space. (ii) In the state $\Omega^{(AB|C)}(\Psi)$ the only virtual system in Alice's side which can be correlated with a virtual system in  Bob's side is the one at the boundary of regions $A$ and $C$, that is the virtual system $a_R$ in Fig.(\ref{fig:MPS}). Similarly, the only virtual system in Bob's side which can be correlated with a virtual system in Alice's side is the one at the boundary of regions $B$ and $C$, that is the virtual system $b_L$ in Fig.(\ref{fig:MPS}).   (iii)  Using the fact that $U_{i}(g)S=S \left[V_{i_L}(g)\otimes {V}^{\ast}_{i_R}(g) \right]$ for all $g\in G$, we find that the charge measurement in the physical space can be interpreted  as a charge measurement in the virtual space, defined by the projectors 
\beq
S^\dag {\Pi}^{(C)}_{\kappa} S= \frac{1}{|G|}\sum_{g\in G} e^{-i\kappa(g)}\bigotimes_{i\in C}  V_{i_L}(g)\otimes {V}^{\ast}_{i_R}(g) \ .
\eeq
Furthermore, because $\forall g\in G: V^\ast_{i_R}(g)\otimes V_{(i+1)_L}(g) |\lambda\rangle =|\lambda\rangle$, the charge for each state $|\lambda\rangle$ is zero. Therefore,  any state $|\lambda\rangle$ whose corresponding pair of virtual systems both live inside region $C$ do not contribute in the total charge  of this region. It follows that the total charge in region $C$ is just determined by  the two unpaired virtual systems at the left and the right boundaries of this region (See Fig.\ref{fig:MPS}).  Putting these facts together, we can easily show proposition \ref{prop23} (Note that $a$ and $b$ in state $\sigma^{(ab)}$ correspond, respectively, to the virtual systems $a_R$ and $b_L$ in Fig.\ref{fig:MPS}).

Therefore, because states  $\sigma^{(ab)}$ and $\Omega^{(AB|C)}(\Psi)$ have the same entanglement properties, to prove theorem \ref{Thm} we can focus on the entanglement properties of state $\sigma^{(ab)}$.

 
\subsubsection{Proof of\  $|\Theta\rangle \xrightarrow{\text{LOCC}} \Omega^{(AB|C)}({\Psi})$
}
It follows from the above proposition that, to prove $|\Theta\rangle \xrightarrow{\text{LOCC}} \Omega^{(AB|C)}({\Psi})$ it suffices to show  $|\Theta\rangle \xrightarrow{\text{LOCC}} \sigma^{(ab)}$. In other words, we need to show that by consuming a copy of the maximally entangled state  $|\Theta\rangle$ one can prepare a copy of state $\sigma^{(ab)}$.

To show this first note that state $\sigma^{(ab)}$ can be generated  in the following way: Alice locally prepares state  $|\lambda\rangle_{aa'}$,  and Bob prepares $|\lambda\rangle_{b'b}$, where
\bes
\begin{align}
(V^\ast_a(g)\otimes V_{a'}(g))|\lambda\rangle_{aa'}&=|\lambda\rangle_{aa'}\ ,\\ 
(V^\ast_{b'}(g)\otimes V_{b}(g))|\lambda\rangle_{b'b}&=|\lambda\rangle_{b'b}\ ,
\end{align}
\ees
for all $g\in G$.  Then, they perform the projective  measurement  $\{\tilde{\Pi}^{({a'b'})}_{\kappa}=\frac{1}{|G|}\sum_{g\in G} e^{-i\kappa(g)}\  V_{a'}(g) \otimes V^\ast_{b'}(g):\kappa\in Q\}$ on systems $a'$ and $b'$, and keep the outcome of the measurement in register $K_C$. In other words, they measure the total charge in the systems ${a'b'}$. Clearly the resulting state will be state $\sigma^{(ab)}$. Therefore, to prove statement $|\Theta\rangle \xrightarrow{\text{LOCC}} \sigma^{(ab)}$, it suffices to show the following
\begin{proposition}\label{proptr}
The projective charge measurement  $\{\tilde{\Pi}^{({a'b'})}_{\kappa}:\kappa\in Q\}$, 
can be implemented via LOCC, by consuming  a maximally entangled state of a pair of $d_{[\omega]}$- dimensional systems, where $d_{[\omega]}$ is the dimension of the irreps in the cohomology class of representation $g\rightarrow V(g) $.
\end{proposition}
To prove this, we assume Alice and Bob are given a pair of $d_{[\omega]}$-dimensional ancillary systems $Z_a$ and $Z_b$, which are prepared in the maximally entangled state $|\Theta\rangle_{Z_aZ_b}\equiv \frac{1}{\sqrt{d_{[\omega]}}}\sum_{k=1}^{d_{[\omega]}} |k\rangle_{Z_a}|k\rangle_{Z_b}\ $.
 Let $g\rightarrow u(g)$ be an irreducible representation of group $G$ in the same cohomology class that the representation $g\rightarrow V(g)$ belongs to it. From lemma \ref{lem-new-dim} we know that all the irreps in the same cohomology class have the same dimension, and therefore the dimension of this representation is $d_{[\omega]}$. Also, note that state $|\Theta\rangle_{Z_aZ_b}$ satisfies 
\beq\label{zero-charge}
[u^\ast(g)\otimes  u(g)]|\Theta\rangle_{Z_aZ_b}=|\Theta\rangle_{Z_aZ_b}\ .
\eeq  
This means that if we assume the representation of symmetry on system $Z_a$ is $g\rightarrow u^\ast(g)$  and on system $Z_b$ is $g\rightarrow u(g)$, then state  $|\Theta\rangle_{Z_aZ_b}$ has charge zero. 

\begin{figure} [h]
\begin{center}
\includegraphics[scale=.29]{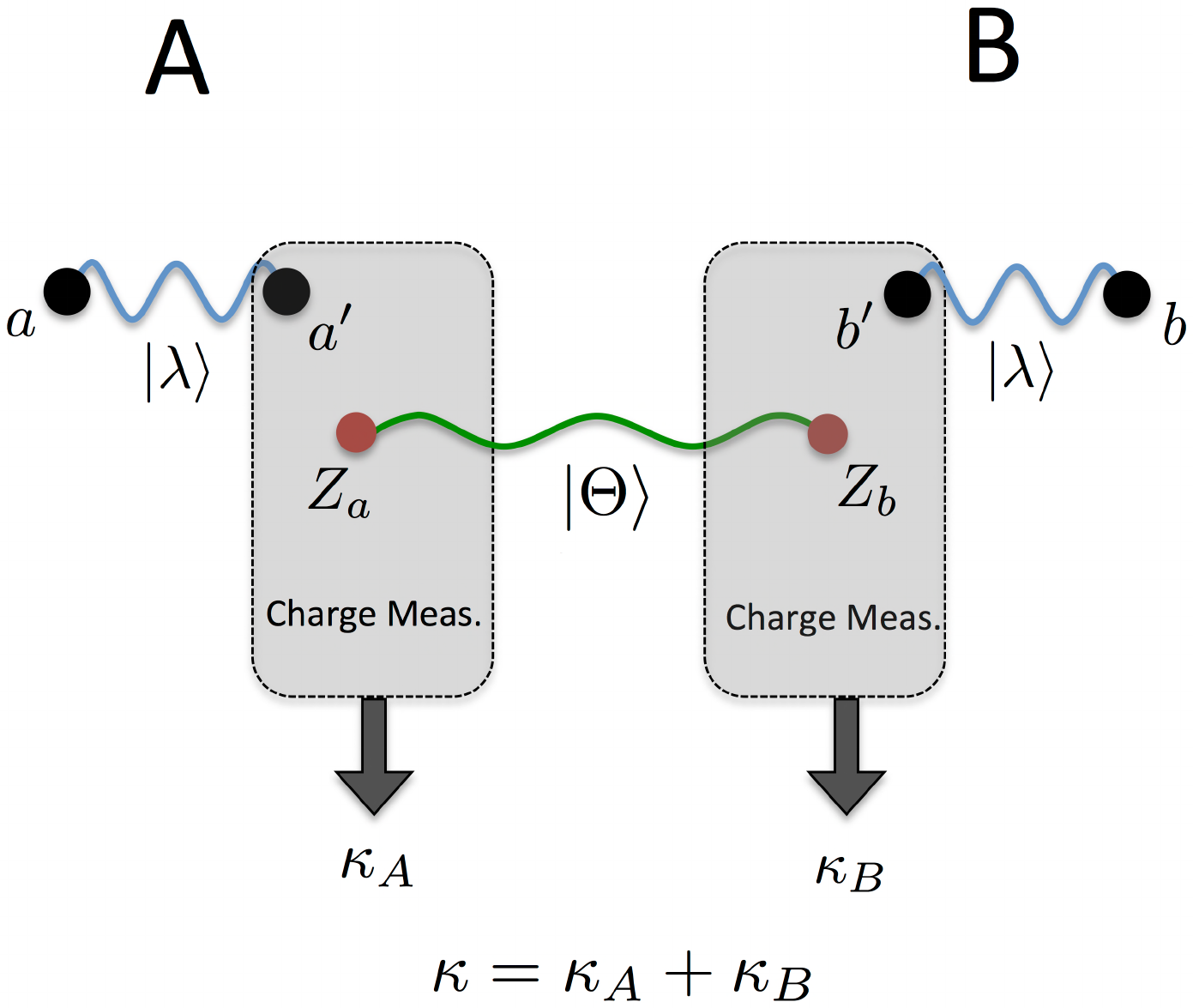}
\caption{By consuming the maximally entangled state $
|\Theta\rangle_{Z_aZ_b}$, Alice and Bob can measure the total charge in systems $a'$ and $b'$, and thereby prepare state $\sigma^{(ab)}$, which is equivalent to state $\Omega^{(AB|C)}(\Psi)$ up to local operations. The idea is  that by adding systems $Z_a$ to the Alice's side and $Z_b$ to the Bob's side,   the representations of symmetry $G$ on both systems $a'Z_a $ and $b'Z_b$ become non-projective. Therefore, Alice and Bob  can locally measure the charges in $a'Z_a $ and   $b'Z_b$. But, because the total charge in  $
|\Theta\rangle_{Z_aZ_b}$ is zero,  the total charge in $a'b'$ is equal to the sum of the charges in $a'Z_a $  and   $b'Z_b$, denoted by  $\kappa_A$  and $\kappa_B$ in the above figure.} \label{fig:Meas}
\end{center}
\end{figure}

It follows that  to perform the charge measurement $\{\tilde{\Pi}^{({a'b'})}_{\kappa}:\kappa\in Q\}$ on systems ${a}'$ and ${b}'$,  Alice and Bob can perform the local charge measurements on systems ${a}' Z_a$ and ${b}' Z_b$ (See Fig.\ref{fig:Meas}). More precisely, Alice performs the measurement corresponding to the projectors
\beq
\tilde{\Pi}^{({a}'Z_a)}_{\kappa}=\frac{1}{|G|}\sum_{g\in G} e^{-i\kappa(g)}\  V_{a'}(g) \otimes u^\ast(g) \ : \kappa\in Q \ ,
\eeq 
on systems ${a}'$ and $Z_a$ and obtains charge $\kappa_A$. Similarly, Bob performs the local projective measurement corresponding to the projectors
\beq
\tilde{\Pi}^{({b'} {Z}_b)}_{\kappa}=\frac{1}{|G|}\sum_{g\in G} e^{-i\kappa(g)}\  V^\ast_{b'}(g) \otimes u(g)\ : \kappa\in Q \ ,
\eeq 
on systems ${b'}$ and $Z_b$, and obtains charge $\kappa_B$. Finally, they add these charges together, and obtain the total charge $\kappa=\kappa_A+\kappa_B$. 

Then, using the fact that state $|\Theta\rangle_{Z_aZ_b}$ has charge zero with respect to the representation $g\rightarrow u^\ast(g)\otimes  u(g)$ (See Eq.(\ref{zero-charge})), we find that adding systems $Z_a$ and $Z_b$ in state $|\Theta\rangle_{Z_aZ_b}$ does not change the total charge. Therefore, the charge $\kappa=\kappa_A+\kappa_B$ is equal to the total charge in systems $a'$ and $b'$.   

This proves proposition \ref{proptr}, and completes the proof of $|\Theta\rangle \xrightarrow{\text{LOCC}} \sigma^{(ab)}$, which by proposition \ref{prop23} implies  $|\Theta\rangle \xrightarrow{\text{LOCC}} \Omega^{(AB|C)}({\Psi})$.

\subsubsection{Proof of $ \Omega^{(AB|C)}({\Psi})\xrightarrow{\text{LOCC}}|\Theta\rangle $
}

In the following, similar to the approach we used in the previous section,  we prove $ \sigma^{(ab)}\xrightarrow{\text{LOCC}}|\Theta\rangle $, and this together with proposition \ref{prop23} imply  $ \Omega^{(AB|C)}({\Psi})\xrightarrow{\text{LOCC}}|\Theta\rangle $.

First, we prove  $ \sigma^{(ab)}\xrightarrow{\text{LOCC}}|\Theta\rangle $, for the special case where the projective representation $g\rightarrow V(g)$ of group $G$ on the virtual systems is an irreducible representation, and then explain how this argument can be extended to the general case. The proof is based on the following lemma
 \begin{lemma}\label{lem11}
Let $g\rightarrow u(g)$ be an irreducible projective representation of an Abelian group $G$. Then, for any 1-dimensional representation $g\rightarrow e^{i\kappa(g)}$ of group $G$, $\frac{1}{|G|}\sum_{g} e^{-i\kappa(g)} [u(g)\otimes  u^\ast(g)]$ is either zero, or a rank one projector $|\Theta(\kappa)\rangle\langle\Theta(\kappa)|$.  Furthermore, $|\Theta(\kappa)\rangle$ is a maximally entangled state.
\end{lemma}
This lemma basically means that for the composite system with represdentation $g\rightarrow u(g)\otimes u^\ast(g)$, any state with a definite charge is a maximllay entangled state.

This lemma follows from the fact that $g\rightarrow  u(g)\otimes  u^\ast(g)$  is a (linear) unitary representation of group $G$, and since the group is Abelian any such representation can be decomposed to 1-dimensional irreps as
\beq
[u(g)\otimes  u^\ast(g)] |\Theta(\kappa)\rangle=e^{i\kappa(g)}  |\Theta(\kappa)\rangle\ ,\ \ \ \forall g\in G
\eeq  
where  $g\rightarrow e^{i\kappa(g)}$ is a 1-dimensional representation of $G$. Then, the fact that $g\rightarrow u(g)$  is an irreducible representation, together with the Schur's lemma implies that each state $|\Theta(\kappa)\rangle$ should be a maximally entangled state. It follows that in  the decomposition of the representation $g\rightarrow u(g)\otimes u^\ast(g)$ to irrpes, each one-dimensional representation  $g\rightarrow e^{i\kappa(g)}$  can show up, at most, once.

From this lemma we find that if the representation $g\rightarrow V(g)$ of symmetry $G$ on the virtual systems is the irreducible representation $g\rightarrow u(g)$,  then
 \begin{align} \label{sre}
\sigma^{(ab)}= \sum_\kappa  p_\kappa |\kappa\rangle\langle\kappa|^{(K_C)}\otimes  |\Theta(-\kappa)\rangle\langle\Theta(-\kappa)|_{{ab}} .
\end{align}
This is true because if register $K_C$  has value $\kappa$ then this means that the charge measurement on ${a'b'}$ has projected them to the charge sector with charge $\kappa$. But, because states $|\lambda\rangle_{aa'}$ and $|\lambda\rangle_{b'b}$ both have charge zero, the initial total charge of systems $ab{a'b'}$ is zero.  This means that if the systems $a'b'$ are projected to state with charge $\kappa$ then the systems  ${ab}$ should be projected to state with the total charge $-\kappa$. Then, by lemma \ref{lem11} we know that there is a unique state of ${ab}$ with charge $-\kappa$, namely state $|\Theta(-\kappa)\rangle$. Furthermore, from this lemma we know that state $ |\Theta(-\kappa)\rangle_{{ab}} $ is a maximally  entangled state. Then, using the fact  all maximally entangle states can be transformed to each other via local unitaries, we conclude that the entanglement of state $\sigma^{(ab)}$ in Eq.(\ref{sre}) is equal to the  entanglement of state $  |\Theta\rangle$. This proves $ \sigma^{(ab)}\xrightarrow{\text{LOCC}}|\Theta\rangle $ for the special case where $g\rightarrow V(g)$ is an irreducible projective representation.

To prove the result for the general case where the representation $g\rightarrow V(g)$ is not irreducible, we first recall that according to lemma \ref{lem-new-dim},  all the irreducible projective representations of an Abelian group in the same cohomology class are equivalent up to a unitary and a phase. This result implies that any projective unitary representation $g\rightarrow V(g)$ of an Abelian group $G$,  induces a tensor product decomposition of the Hilbert space  as  $\mathcal{H}\equiv \mathcal{M}\otimes\mathcal{N}$, where the subsystem $\mathcal{M}$ corresponds to $g\rightarrow u(g)$, an irreducible projective representation in the same equivalence class of the representation $g\rightarrow V(g)$, and $\mathcal{N}$ is the corresponding multiplicity subsystem.  In other words, there exists a unitary $W$ such that 
\beq
\forall g\in G:\  \ WV(g) W^\dag=u(g)\otimes \sum_r e^{i\theta_r(g)} |r\rangle\langle r| \ ,
\eeq
 where $g\rightarrow u(g)$ acts irreducibly on $\mathcal{M}$, and   $\{|r\rangle\} $  is a basis for the  subsystem $\mathcal{N}$, and $e^{i\theta_r(g)}$ are arbitrary phases. 
    
Using this observation, we can extend the above proof to the case where the representation $g\rightarrow V(g)$ is not irreducible: Because the initial states $|\lambda\rangle_{aa'}$ and $|\lambda\rangle_{b'b}$ both have charge zero, if the charge measurement on systems ${a'b'}$ projects them to the subspace with charge $\kappa$, then systems $ab$ should be in charge $-\kappa$ sector. Then, if Alice and Bob perform  projective measurements on multiplicity subsystems $\mathcal{N}$ of the virtual systems $a$ and $b$, they will find the total charge of the irreducible subsystems $\mathcal{M}$ of systems $a$ and $b$. But, from lemma \ref{lem11} we know that any state of $ab$ which has a definite charge should be a maximally entangled state. This proves that $ \sigma^{(ab)}\xrightarrow{\text{LOCC}}|\Theta\rangle $, and completes the proof of theorem.

\section{SPT-entanglement in arbitrary dimension}\label{Sec:D}

In this section we study some general properties of SPT-Entanglement, which hold in arbitrary dimension.  In particular, we study how SPT-Entanglement changes under the effect of low-depth symmetric circuits.

\subsection{Monotonicity of SPT-Entanglement with the size of regions $A$ and $B$}\label{Sec:monoton}
\begin{figure} [h]
\begin{center}
\includegraphics[scale=.2]{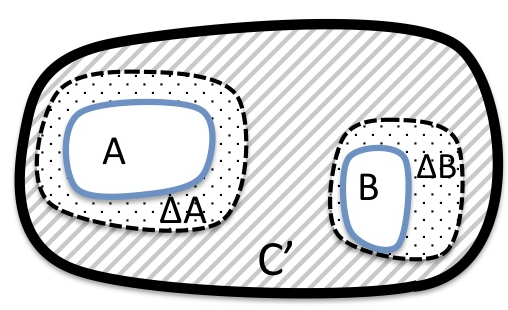}
\caption{Region $A'$ is partitioned to region $A$ and region $\Delta A$. Similarly, region $B'$ is partitioned to region $B$ and region $\Delta B$. Region ${C}'\equiv C\setminus \left(A'\cup B'\right)$ is the subregion of $C$ surrounded by $A'$, $B'$ and the boundaries of the system. } \label{fig:Extended}
\end{center}
\end{figure}

Our first result states that as we make regions $A$ and $B$ larger, SPT-entanglement either remains constant or increases. More formally,
\begin{proposition}\label{momont:Thm}
Let $A'$ and $B'$ be a pair of non-overlapping regions  that contain $A$ and $B$ respectively, that is $A\subseteq A'$ and $B\subseteq B'$. Let $C$ be a region surrounded by $A$, $B$ and the boundaries of the system, and ${C}'\equiv C\setminus \left(A'\cup B'\right)$ be the subregion of $C$ surrounded by $A'$, $B'$ and the boundaries of the system (See Fig.\ref{fig:Extended}). Then, 
\beq
\Omega^{\left(A'B'|C'\right)}(\rho)\xrightarrow{\text{LOCC}}\Omega^{(AB|C)}(\rho) ,
\eeq
Therefore, for any measure of entanglement $E$, it holds that 
$ E(\Omega^{(AB|C)}(\rho)) \le E(\Omega^{\left(A'B'|C'\right)}(\rho))$ .
\end{proposition}
If we choose negativity as a measure of entanglement then this implies 
\beq
\sum_{\kappa}\left\| \left[\text{tr}_{\overline{AB}}(\Pi^{(C)}_{\kappa} \rho)\right]^{\textbf{T}_{A}}   \right\|_{1}\ \le \sum_{\kappa}\left\| \left[\text{tr}_{\overline{A'B'}}(\Pi^{(C')}_{\kappa} \rho)\right]^{\textbf{T}_{A}}   \right\|_{1}\ .
\eeq
Note that here we are looking at the bipartite entanglement of state 
\beq
\Omega^{\left(A'B'|C'\right)}(\rho)=\sum_\kappa |\kappa\rangle\langle\kappa|^{(K_{C'})}\otimes \text{tr}_{\overline{A'B'}}(\Pi^{(C')}_{\kappa} \rho) \ ,
\eeq
where the systems $K_{C'}$, $A'$ and $B'$ are partitioned as $A'|K_{C'}B'$ (or equivalently $A'K_{C'}|B'$). In other words, all the sites in region $A'$ are given to Alice,  and all the sites in region $B'$ are given to Bob, and the classical register $K_{C'}$, which keeps the information about the charge in region $C'$,  is given to either Alice or Bob.

The proof of proposition  \ref{momont:Thm} is simple: Let $\Delta A=A'\setminus A$ and $\Delta B=B'\setminus B$ be, respectively,  the subset of region $A'$ which is not in $A$, and the subset of region $B'$ which is not in $B$ (See Fig. \ref{fig:Extended}). Alice and Bob, who are given all the sites in regions $A'$ and $B'$ respectively, can measure the  charge of the sites in regions $\Delta A$ and   $\Delta B$. Then, because the charge is Abelian, the total charge in region $C$ is 
\beq
\kappa_C=\kappa_{C'}+\kappa_{\Delta A}+\kappa_{\Delta B}\ \ \ (\text{mod}\ 2\pi) ,
\eeq
where $\kappa_{C'}$ is the charge in regions $C'$, $\kappa_{\Delta A}$ is the charge in region $\Delta A$,  and  $\kappa_{\Delta B}$ is the charge in region $\Delta B$. Therefore, adding  the charges $\kappa_{\Delta A}$ and $\kappa_{\Delta B}$ to the charge in region $C'$, whose value is written in the register $K_{C'}$,  they find the total charge in region $C$, and they  update the value of this register, i.e. they implement the following transformation
\beq
|\kappa_{C'}\rangle^{(K_{C'})}  \longrightarrow |\kappa_{C'}+\kappa_{\Delta A}+\kappa_{\Delta B}\  (\text{mod}\ 2\pi) \rangle^{(K_{C})}  . 
\eeq
 Finally, by tracing over the sites in regions $\Delta A$ and $\Delta B$ they can transform state $\Omega^{\left(A'B'|C'\right)}(\rho)$ to state $\Omega^{\left(AB|C\right)}(\rho)$.  Note that to find the total charge in region $C$ and update the classical register, Alice and Bob need to have classical communication with each other.

\subsection{Effect of low-depth symmetric circuits on SPT-Entanglement}\label{Sec:effect}
Next, we consider the effect of low-depth symmetric circuits on SPT-Entanglement, and show that under such transformations SPT-Entanglement cannot change drastically. 



 Let $A$ and $B$ be two non-overlapping regions of the system and $C$ be a region surrounded by $A$, $B$ and the boundaries of the system. In the following, to simplify the presentation, we assume these three regions cover the  entire system, and there is no site outside these regions.  For any region $X$ let $X_{R}$ be the set of sites in the ball of radius $R$ from $X$, that is the set of sites whose distance from $X$ is less than or equal to $R$. Similarly, for any region $X$ let $X_{-R}$ be a subset of $X$ whose distance from the complement of $X$ is larger than $R$. Note that region $C_{-R}$  is surrounded by regions $A_R$ and $B_R$, and the boundaries of the system, and similarly  region $C_{R}$  is surrounded by regions $A_{-R}$ and $B_{-R}$, and the boundaries of the system. Therefore the sets $\{A,B,C\}$, $\{A_R,B_R,C_{-R}\}$ and $\{A_{-R},B_{-R},C_{R}\}$ define three different ways of  partitioning  the sites of systems to three regions.

 \begin{theorem}\label{Thm:LOCC-transmain}
Let $V$ be a symmetric circuit with range $R$ which is bounded by $R<\text{dist}(A,B)/2$.  Then, the following transformations can be implemented via LOCC, 
\begin{align}
\Omega^{(A_{R}B_{R}|C_{-R})}(\rho) \xrightarrow{} \Omega^{\left(AB|{C}\right)}(V \rho V^{\dag})&\xrightarrow{}\Omega^{(A_{-R}B_{-R}|C_R)}(\rho) \ .
 \end{align}
That is there exist LOCC transformations which transform $\Omega^{(A_{R}B_{R}|C_{-R})}(\rho)$ to $\Omega^{\left(AB|{C}\right)}(V \rho V^{\dag})$, and the latter state to $\Omega^{(A_{-R}B_{-R}|C_{R})}(\rho)$. Therefore, for any measure of entanglement $E$, 
\begin{align}
E(\Omega^{(A_{-R}B_{-R}|C_R)}(\rho))&\le  E(  \Omega^{\left(AB|C\right)}(V \rho V^{\dag}) )\nonumber\\ &\le E(\Omega^{(A_{R}B_{R}|C_{-R})}(\rho) )\ .
\end{align}
\end{theorem}


In words, this inequality means that the SPT-Entanglement of state  $V\rho V^\dag$ for regions $A$ and $B$ is lower bounded by the SPT-Entanglement of $\rho$ between $A_{-R}$ and $B_{-R}$, and upper bounded by  the SPT-Entanglement of $\rho$ between $A_{R}$ and $B_{R}$. 

In Sec.(\ref{Sec:main}) we sketched a proof of this theorem. Here, we present another proof, which is based on a slightly different point of view.

\begin{proof}
The key point in the argument is that because $V$ is a low-depth circuit with range $R$, having access to all the sites in the extended regions  $A_R$ and $B_R$, Alice and Bob can apply all the local unitary gates in this circuit which are in the light cones of region $A$ and region $B$ (See Fig.(\ref{fig:low-depth})).  Call these local symmetric unitaries which act non-trivially on $A_R$ and $B_R$, $V_{A_R}$  and $V_{B_R}$, respectively. Then, for any operator $X$, 
\beq\label{eq_reg}
\Tr_{\overline{AB}}\big(V X V^\dag\big)= \Tr_{\overline{AB}}\big([V_{A_R}\otimes V_{B_R}]\ X \ [V^\dag_{A_R}\otimes V^\dag_{B_R}]\big)\ ,
\eeq
where the partial traces is over all sites in the system except those which are in $A$ and $B$. Note that this equation is simply a consequence of the fact that $V$ is low-depth. In particular, applying this equation for $X=\rho$ we find
\beq
\Tr_{\overline{AB}}\big(V \rho V^\dag\big)= \Tr_{\overline{AB}}\big([V_{A_R}\otimes V_{B_R}]\ \rho \ [V^\dag_{A_R}\otimes V^\dag_{B_R}]\big)\ .
\eeq
Next, using the fact that both unitaries $V$ and $V_{A_R}\otimes V_{B_R}$ are symmetric, and hence preserve the total charge, together wit the fact that the reduced state of regions $A$ and $B$ is the same for  states $V\rho V^\dag$ and  $(V_{A_R}\otimes V_{B_R})\rho (V^\dag_{A_R}\otimes V^\dag_{B_R})$, we can easily show that the charge in region $C$ is also the same for these two states (Note that the charge in region $C$ can be thought as the total charge in the system minus the charge in regions $A$ and $B$). In other words, applying Eq.(\ref{eq_reg}) for $X=\Pi_\kappa\rho$, where $\Pi_\kappa$ is the projector to the subspace with charge $\kappa$ in the system, and using $[\Pi_\kappa, V]=[\Pi_\kappa, V_{A_R}\otimes V_{B_R}]=0$, we can easily show that
\beq
\Tr_{\overline{AB}}\big(\Pi_\kappa^{(C)} V\rho V^\dag\big)= \Tr_{\overline{AB}}\big(\Pi_\kappa^{(C)} [V_{A_R}\otimes V_{B_R}]\ \rho \ [V^\dag_{A_R}\otimes V^\dag_{B_R}]\big)\ ,
\eeq
where $\Pi_\kappa^{(C)}$ is the projector with charge $\kappa$ in region $C$.

Using this observation we can easily find a LOCC protocol which transforms state 
 \beq
\Omega^{\left(A_{R}B_{R}|C_{-R}\right)}(\rho)=\sum_{\kappa}  |\kappa\rangle\langle\kappa|^{K} \otimes \Tr_{\overline{A_{R}B_{R}}}(\rho \Pi^{(C_{-R})}_\kappa) \ ,
\eeq
to state 
\beq
\Omega^{\left(AB|C\right)}(V \rho V^\dag)=\sum_{\kappa}  |\kappa\rangle\langle\kappa|^{K} \otimes \Tr_{\overline{A}\overline{B}}( \Pi^{(C)}_\kappa V \rho V^\dag) \ .
 \eeq
The protocol includes the following steps:  (i) Alice and Bob apply the local symmetric unitaries $V_{A_R}$ and $V_{B_R}$ defined above on the sites in regions $A_R$ and $B_R$.  (ii) They measure the charges in regions $\Delta A\equiv A_R\setminus A$  and $\Delta B\equiv B_R\setminus B$. (iii) They add the charges obtained in these regions to the charge in region $C_{-R}$ which is written in register $K$, to find the total charge in region $C$ for state $\rho$, and write down this charge in register $K$. (iv) They discard all the sites in regions $\Delta A$, and $\Delta B$. 

One can easily show that this protocol implements the following transformation 
\beq
\Omega^{\left(A_RB_R|{C}_{-R}\right)}(\rho)\xrightarrow{\text{LOCC}}\Omega^{(AB|C)}(V\rho V^\dag)\ . 
\eeq
Next, to prove the rest of the theorem, we use the fact that  if $V$ is a symmetric circuit with range $R$, then $V^\dag$ is also a symmetric circuit with range $R$. Then, applying the above result with $V^\dag$ instead of $V$, and $V\rho V^\dag$ instead of $\rho$, we can easily show that 
\beq
\Omega^{\left(AB|{C}\right)}(V\rho V^\dag)\xrightarrow{\text{LOCC}}\Omega^{(A_{-R}B_{-R}|C_R)}(\rho)\ . 
\eeq
This completes the proof.
\end{proof}

\subsubsection{SPT-Entanglement vanishes in the trivial phase}
Using this result, we can easily show that SPT-Entanglement should vanish for all states in the trivial phase. 
\begin{corollary}\label{Thm0}
Suppose there exists  a symmetric circuit with range $R$, which transforms state $\rho$ to a product state. Then, the SPT-Entanglement of state $\rho$ between any two regions $A$ and $B$ with distance more than $2R$ is zero, i.e.  for any measure of entanglement $E$, it holds that $\text{dist}(A,B)>2R$ implies $E\left( \Omega^{(AB|C)}(\rho)\right) =0.$
\end{corollary}


\subsubsection{SPT-Entanglement is universal in all phases}

Corollary  \ref{Thm0} implies that in the trivial phase SPT-Entanglement remains constant throughout the phase. This result was a corollary of theorem \ref{Thm:LOCC-transmain} which shows that the effect of symmetric low-depth circuits on SPT-Entanglement can be bounded by looking to the small deformations of the boundaries of the regions $A$ and $B$.  

Next, we consider the consequences of theorem  \ref{Thm:LOCC-transmain} for SPT-Entanglement in a general, non-trivial  SPT phase.  In particular, we show that, assuming the SPT-Entanglement remains unchanged under small deformations of the boundaries of regions $A$ and $B$, then this theorem implies that the SPT-Entanglement should remain constant under the effect of symmetric low-depth circuits. That is the SPT-Entanglement should be constant throughout a phase.



To explain the assumption, first recall that as we showed in proposition \ref{momont:Thm} by making regions $A$ and $B$ larger SPT-Entanglement cannot decrease, i.e.
\begin{equation}\label{LOCC-trans}
A\subseteq A',B\subseteq B'\ \Rightarrow \ \Omega^{\left(A'B'|{C'}\right)}(\rho)\xrightarrow{\text{LOCC}}\Omega^{(AB|C)}(\rho)\ .
\end{equation}
The above relation holds for arbitrary state $\rho$ and arbitrary regions $A$ and $B$. For a general state,  the SPT-Entanglement  between the larger regions $A'$ and $B'$ is larger than the SPT-Entanglement between the smaller regions $A$ and $B$, and hence this transformation is not reversible via LOCC. However, assuming  that the system is sufficiently homogenous and regions $A$ and $B$ are sufficiently large and far from each other compared to the correlation length, one expects that the SPT-Entanglement should saturate, and therefore a small increase in the sizes of regions $A$ and $B$ does not increase it anymore. In other words, it seems natural to assume that in this limit the inverse transformation should also be possible via LOCC, i.e $\Omega^{(AB|C)}(\rho)\xrightarrow{\text{LOCC}} \Omega^{\left(A'B'|{C'}\right)}(\rho)$. Note that, as we saw in the previous section, our results based on MPS representation of SPT phases proves the validity of this assumption in the case of  1-dimensional systems. However, the validity of this assumption in the higher dimensions is still an open problem.

In the following, to simplify the presentation, we assume regions $A$, $B$ and $C$ cover the entire system.

For any region $X$ let $X_{+R}$ be the set of sites in the ball of radius $R$ from $X$, that is the set of sites whose distance from $X$ is less than or equal to $R$. Similarly, for any region $X$ let $X_{-R}$ be a subset of $X$ whose distance from the complement of $X$ is larger than $R$.  

Then, using theorem \ref{Thm:LOCC-transmain} we can easily show that
\begin{corollary}\label{Thm1}
Suppose state $\rho$ satisfies the condition
$\Omega^{(A_{-R}B_{-R}|C_{+R})}(\rho)\xrightarrow{\text{LOCC}}\Omega^{\left(A_{+R}B_{+R}|{C}_{-R}\right)}(\rho)$. That is the SPT-Entanglement for smaller regions $A_{-R}$ and $B_{-R}$ is the same as the SPT-Entanglement for the larger regions   $A_{+R}$ and $B_{+R}$. Then, for any symmetric circuit $V$  with range $R$,  the SPT-Entanglement of  $\rho$ and $V\rho V^\dag$ between $A$ and $B$ relative to $C$  are equal, i.e.   $\Omega^{(AB|{C})}({\rho})\xleftrightarrow{\text{LOCC}}\Omega^{(AB|{C})}(V{\rho}V^\dag).$ Therefore, for any measure of entanglement $E(\Omega^{(AB|{C})}({\rho}))=E(\Omega^{(AB|{C})}(V{\rho}V^\dag))$. 
\end{corollary}

\section{Conclusion}
 
 The fact that symmetry-protected topological orders should admit an entanglement-based order parameter measuring the nonlocal entanglement between the edge degrees of freedom, seems natural. In this work, we provided an explicit example of such order parameters. Remarkably, this order parameter turns out to be closely related to the string order parameters, via Fourier transform.

Another appealing  aspect of this work is to show how the resource theoretic point of view to entanglement could be useful in the context of many-body systems. If one tries to formulate  the notion of SPT-Entanglement in terms of a particular measure of entanglement, such as negativity, then understanding and proving the properties of SPT-Entanglement will be much harder. Furthermore, using the resource theoretic point of view to entanglement enabled us to clearly see how the universality of the SPT-entanglement relies on   the defining property of measures of entanglement, namely their monotonicity under classical communication and local operation. Quantum resource theories have recently attracted a lot of attention in the quantum information community, and it is interesting to see if they find other applications in the field of many-body systems.

Another feature of the SPT-Entanglement is that it uses measures of \emph{bipartite} entanglement to capture tripartite correlations in the system, that is correlations between local degrees of freedom in regions $A$ and $B$, and a classical degree of freedom in region $C$, namely the total charge in this region. It is interesting to see if this sort of quantification of tripartite correlations via bipartite entanglement measures have any other applications in many-body systems (See \cite{Isaac, Osborne} for a different approach to tripartite correlations).

Finally, we noted that the identity gate in MBQC  can be interpreted as the charge measurement for the group $\mathbb{Z}_2\times\mathbb{Z}_2$. This simple observation may, to some extent, demystify the robustness of the computational power of 1-dimensional cluster state under symmetric perturbation.

\section{Acknowledgment}
I would like to thank Stephen Bartlett and Paolo Zanardi for reading the manuscript and providing very useful comments. Also, I acknowledge helpful discussions with John Preskill, Lorenzo Venuti, Benny Yoshida, Isaac Kim, Spyridon Michalakis, Steve Flammia, and Jacob Bridgeman. This research was supported by ARO MURI grant W911NF-11-1-0268.

\newpage

\onecolumngrid
\appendix

\section{Projective irreducible representations of Abelian groups (Proof of lemma \ref{lem-new-dim})}
In the following we repeat the statement of  lemma  \ref{lem-new-dim} and prove it:

\noindent\textbf{Lemma}
\emph{Let $g\rightarrow u_\beta(g)$ and $g\rightarrow u_\gamma(g)$  be two finite-dimensional irreducible projective representations of an Abelian group $G$ whose 2-cocyles belong to the same cohomology class. Then, their dimensions is the same. Furthermore there exists a unitary $W$, and a phase $e^{i  r(g)} $ such that  }
\beq\label{eq:fin}
\forall g\in G:\ \ u_\beta(g)=e^{i r(g)} Wu_\gamma(g) W^{\dag}\ .
\eeq

\begin{proof}
Let $\omega(g,h)$ be the 2-cocycle of the representation  $g\rightarrow u_\beta(g)$, i.e.
\beq
u_\beta(g) u_\beta(h)=\omega(g,h) u_\beta(gh) \ .
\eeq
Since the 2-cocycle of the representations  $g\rightarrow u_\gamma(g)$ and $g\rightarrow u_\beta(g)$  belong to the same cohomology class, we  know that there exists a phase $e^{i s(g)}$, such that the 2-cocycle for the representation  $g\rightarrow e^{is(g)} u_\gamma(g)$, is also  $\omega(g,h)$, that is 
\beq
e^{is(g)}u_\gamma(g) e^{is(h)}u_\gamma(h)=\omega(g,h) e^{is(gh)}u_\gamma(gh) \ .
\eeq
Let ${u}_\beta^{\ast}(g)$ be the complex conjugate of ${u}_\beta(g)$. Then, we can easily see that the representation $g\rightarrow u^\ast_\beta(g)\otimes e^{is(g)} u_\gamma(g)$  is a (non-projective) unitary representation of $G$, that is its 2-cocycle is trivial. 

Next, we note that since the group is Abelian all of its non-projective irreducible representations are  1-dimensional, and therefore the representation  $g\rightarrow u^\ast_\beta(g)\otimes e^{is(g)} u_\gamma(g)$ can be decomposed to 1-dimensional  irreducible representations of $G$. Let the normalized vector $|\Theta(\kappa)\rangle$ be a 1-dimensional subspace on which  $g\rightarrow u^\ast_\beta(g)\otimes e^{is(g)} u_\gamma(g)$ acts irreducibly, i.e.
\begin{equation}\label{Eq-app}
\forall g\in G:\ \ \left[u^\ast_\beta(g)\otimes e^{is(g)} u_\gamma(g)\right]|\Theta(\kappa)\rangle=e^{i \kappa(g)}|\Theta(\kappa)\rangle \ ,
\end{equation}
for some 1-dimensional representation $e^{i \kappa(g)}$.  Then, using the fact that the representations $\beta$ and $\gamma$ are irreducible, it follows from the Schur's lemma that $|\Theta(\kappa)\rangle$ should be a maximally entangled state, and its corresponding reduced state on both subsystems should be proportional to the identity operator 
(To see this consider the reduced density operator corresponding to state $|\Theta(\kappa)\rangle$ on one subsystem, and then   use Eq.(\ref{Eq-app}) to show that this reduced state should commute with an irreducible representation. Then, using Schur's lemma we find that this density operator should be proportional to the identity operator, which means state $|\Theta(\kappa)\rangle$  should be maximally entangled.).

This implies that the dimension of the irreducible representations  $\beta$ and $\gamma$ are the same. We denote this dimension by $d$. Furthermore, since $|\Theta(\kappa)\rangle$ is a maximally entangled normalized vector, there exists a unitary $V$ such that 
\beq
|\Theta(\kappa)\rangle=(V\otimes I) |\psi^{(+)}\rangle=(V\otimes I)  \frac{1}{\sqrt{d}} \sum_{i=1}^{d} |ii\rangle\ ,
\eeq
where $|\psi^{(+)}\rangle\equiv\frac{1}{\sqrt{d}} \sum_{i=1}^{d} |ii\rangle$, and $I$ is the identity operator acting on the $d$-dimensional space. Then Eq.(\ref{Eq-app}) implies
\begin{equation}
\left[(V^{\dag}u^\ast_\beta(g)V)\otimes u_\gamma(g)\right]|\psi^{(+)}\rangle= e^{i (\kappa(g)-s(g))}|\psi^{(+)}\rangle\ .
\end{equation}
Next, we use the fact that for any  operator $B$ we have $(I\otimes B)   |\psi^{(+)}\rangle= (B^{\textbf{T}}\otimes I)   |\psi^{(+)}\rangle$, where  $B^{\textbf{T}}$ is the transpose of $B$. It follows that
\begin{equation}
\left[\left( V^{\dag}u^\ast_\beta(g)V  u_\gamma^{T}(g)\right)\otimes I\right] |\psi^{(+)}\rangle= e^{i (\kappa(g)-s(g))} |\psi^{(+)}\rangle\ .
\end{equation}
This implies that $V^{\dag}u^\ast _\beta(g)V  u_\gamma^{T}(g)= e^{i (\kappa(g)-s(g))}I$, and therefore
\begin{equation}
\forall g\in G:\ u_\gamma(g) =e^{i (\kappa(g)-s(g))} V^T u_\beta(g) V^{\ast}\ .
\end{equation}
Choosing $W=V^T$, and $e^{i r(g)}= e^{i (\kappa(g)-s(g))}$ we obtain Eq.(\ref{eq:fin}).
\end{proof}

\end{document}